# Near-unity broadband omnidirectional emissivity via femtosecond laser surface processing

*Andrew Reicks[1], Alfred Tsubaki[1], Mark Anderson[2], Jace Wieseler[2], Larousse Khosravi Khorashad[1], Jeffrey E. Shield[2], George Gogos[2], Dennis Alexander[1], *Christos Argyropoulos[1], and *Craig Zuhlke[1]

[1] Department of Electrical & Computer Engineering, University of Nebraska-Lincoln, 1400 R St. Lincoln, Nebraska 68588, USA

[2] Department of Mechanical & Materials Engineering, University of Nebraska-Lincoln, 1400 R St. Lincoln, Nebraska 68588, USA

## Abstract

It is very challenging to achieve near perfect absorption/emission that is both broadband and omnidirectional while utilizing a scalable fabrication process. Femtosecond laser surface processing is an emerging low-cost and large-scale manufacturing technique used to directly and permanently modify the surface properties of a material. The versatility of this technique to produce tailored surface properties has resulted in a rapidly growing number of applications. Here, we demonstrate near perfect, broadband, omnidirectional emissivity from aluminum surfaces by tuning the laser surface processing parameters including fluence, pulse count, and the ambient gas. Full-wave simulations and experimental results prove that the obtained increase in emissivity is mainly a result of two distinct features produced by femtosecond laser surface processing: the introduction of microscale surface features and the thick oxide layer. This technique leads to functionalized metallic surfaces that are ideal for emerging applications, such as passive radiative cooling and thermal management of spacecraft.

# Introduction

Recently, a substantial amount of research efforts have focused on developing surfaces with high electromagnetic absorption or emission in the infrared (IR) regions of the electromagnetic spectrum with important applications in passive radiative cooling[1–3], thermophotovoltaics[4–6], and thermal management of spacecraft[7–9]. Typically, state-of-the-art surfaces with high electromagnetic absorption/emission can be divided into three categories: coatings and paints, metamaterials, and laser processed surfaces.

Coatings and paints are similar approaches to increasing emissivity; they are utilized to add a layer or layers of a material to obtain surface properties different than those of the substrate. Coatings and paints technologies vary significantly in terms of materials, thickness, number of layers, and application method[9–17]. Coatings are usually designed to utilize the emission properties of the low index material and the high absorption caused by the phonon-polariton resonance of the high index material at IR frequencies[5,11,15,16]. Paints can vary significantly on how the high emissivity response is achieved, however, many are based on organic compounds or oxide nanoparticles[3,4,10,11,18]. Coatings and paints have a number of advantages that have led to their widespread usage[10,14,15], including affordability and the relative ease in which they can be applied to nearly any material. Additionally, several offer tunable absorption over most of the IR spectrum, however this is typically narrowband[12,13]. Coatings and paints have similar disadvantages including being prone to delamination and easy degradation with time, especially in harsh environments such as space[5,7,19]. Since they are relatively smooth, most suffer from high angular sensitivity[9,14]. Additionally, most high-emissivity coatings or paints require time to fully cure before they can be used, usually up to seven days[18], and utilize toxic materials[3,4,16].

Recently, wide-angle, high absorption/emission responses have been demonstrated with metallic (plasmonic)[20] or dielectric[17,21,22] metamaterial structures. It has also been demonstrated that metallic gratings can be used to produce near perfect emissivity at a chosen wavelength and angle[23]. Similarly, tapered and elongated gratings can offer near perfect absorption across several angles in the visible spectrum[24,25]. Narrowband absorption in the IR spectrum has also been demonstrated by using different surface shapes, such as crosses, circles, and squares[26–28]. Using other shapes like "trapezoidal ridges" offers absorption over a broader spectral band, and the use of grids offers high absorption at a wide range of angles[29]. However, all of these structures result in enhanced absorption/emission over a narrow spectral band, typically over just a few micrometers. In addition, their response is always angle-dependent, and they do not operate as perfect absorbers at grazing angles. Recently, theoretical works have demonstrated tunable, near-perfect, wide-angle absorption over a variety of wavelength ranges in the IR spectrum by using alternating metal-dielectric layers and metamaterials with different surface shapes such as columns, pyramids, or trapezoidal structures[30–32]. Nevertheless, the experimental verification of these structures is still elusive, mainly due to the complexity of the required niche fabrication processes. Moreover, most applications of high emissivity surfaces require large area inexpensive absorbers, while most metamaterial structures can currently only be produced over extremely small areas using costly high accuracy lithographic techniques. In addition, the perfectly periodic nature required of these metamaterials is prone to fabrication imperfections, so typically high emissivity is obtained for only a narrow spectral range as compared to the broadband results that have been demonstrated using coatings or laser processing.

Many previous studies have demonstrated that laser processing can be used to modify how surfaces reflect, absorb, or emit light[9,15], including large increases in broadband absorption/emission on surfaces processed using short pulsed lasers. The surfaces are generally produced either by using femtosecond laser surface processing (FLSP) to form quasi-periodic self-organizing microstructures[33–37] or by directly writing pattens such as a grid or array of holes onto the surface[38–41]. However, none of these papers report surfaces with emissivity that is near perfect, omnidirectional, and broadband. Broadband moderate absorption values have been demonstrated over a wide spectral range from 0.3 to 50 μm on aluminum processed using a femtosecond laser at relatively high fluence (13.5 J/cm$^2$) to create quasi-periodic surface structures[33]. However, that work was focused on the broadband absorption of the surfaces and no work was completed to fine-tune the surfaces to maximize the emissivity. Periodic submicron ripples produced using low fluence values, known as laser induced period surface structures (LIPSS), can be used to produce high absorption in narrow bands that are tunable over a wide spectral range from 250 nm to 300 μm on aluminum[36], very similar to the results demonstrated for metamaterial structures. Research on LIPSS has been expanded up to fluence values of 2.4 J/cm$^2$ to include microscale structures on aluminum with similar narrowband absorption obtained in the limited spectral band of 0.4 to 1 μm [35]. Another study of laser processed surfaces demonstrated moderately high absorption in the visible spectrum based on aluminum using quasi-periodic self-organized microstructures, in addition to increased absorption from 2.5 to 15 μm on titanium and stainless steel, but the emissivity was only measured for angles of 10, 40, and 60 degrees from the surface normal[37]. While these studies hypothesize that roughness and surface chemistry are possible causes for the increase in absorption on aluminum, none address these important issues experimentally. Other researchers have produced similar results with moderately high absorption at normal incidence over the visible spectrum and into the IR, out to 2.5 µm, by directly laser writing a grid pattern on copper[39,40], aluminum[37], and silicon[38]. However, these direct laser writing methods do not lead to dynamic structures in terms of high aspect ratios and roughness that can be produced by the FLSP process. As a result, these surfaces fail to produce a broadband near perfect omnidirectional response. Furthermore, the direct laser writing methods likely do not result in a thick oxide layer like the surfaces reported in this paper, because with direct writing the laser only interacts with specific parts of the surface (e.g., area within the channels). In addition, all the previous relevant studies do not include detailed materials science analysis to quantify the oxide layer thickness and formation dynamics. Many previous works reported in the literature have indicated three possible causes leading to the increase in broadband absorption of the laser processed material: addition of micro/nanoscale surface structures[37,39,40,42], changes in chemistry[38,42,43], and the effect of impedance matching[41,43]. Many of these studies examine only a single cause and most do not address aluminum. Understanding the role that both the oxide layer and surface microstructure plays in modifying surface properties requires a complete analysis of the laser processed surfaces including subsurface chemical and microstructure analysis, which are performed in this work.

FLSP is an emerging advanced manufacturing technique that can be used to directly alter the properties of a surface. With FLSP, permanent multiscale surface features are produced that are typically characterized by microscale mounds, or pyramidal structures, covered by a layer of redeposited nanoparticles[44–47]. The resulting micro and nanoscale roughness, along with modified surface chemistry and subsurface microstructure, accounts for the unique properties attributed to these surfaces. These features form through a unique combination of ablation, redeposition, melting, fluid flow and resolidification[48]. The surface morphology and chemistry can be directly controlled by processing parameters such as fluence, the number of pulses applied, and the atmospheric environment present when processing the surface[49,50]. The versatility of FLSP for producing tailored surface properties results in a wide range of applications, including improved anti-bacterial response[51–53], modified wettability[54,55], enhanced heat transfer properties[56,57], and tunable electromagnetic response[33,35,36].

In this work, we theoretically and experimentally demonstrate, for the first-time, near perfect hemispherical emissivity in the spectral range from 7.5 to 14 μm, exhibited by aluminum surfaces processed using FLSP within an air environment. Emissivity was measured over this spectral range because it is an atmospheric window of interest for many thermal management applications and is the range of operation of the thermal camera used for the measurements. The developed FLSP surfaces outperform the emissivity response of all coatings and metamaterial structures presented in the literature. Furthermore, the FLSP technique has many advantages over other surface functionalization techniques: it results in a fully functionalized surface in a single processing step; it is a scalable process; it involves the creation of hierarchical micro and nanoscale surface features composed of the original material, making the surface highly permanent; it leads to modification of the original surface without the net addition of mass; and it results in a minimized heat affected zone, so the surface can be modified without altering the bulk properties of the material[58]. We use experimental and theoretical insights to prove that both surface oxidation and microscale surface features play key roles in the large emissivity increase. A detailed surface and subsurface analysis of chemistry, porosity, and microstructure enables the complete characterization of the FLSP surfaces and provides inputs to the performed theoretical modeling of light scattering from these surfaces. The laser processed surfaces produced are ideal candidates to be used as a permanent solution to achieve passive radiative cooling of large area metallic surfaces, thermophotovoltaics, thermal management in spacecrafts, and energy absorption for laser power beaming or stealth technologies.

## Results

**Broadband and omnidirectional emissivity response.** We demonstrate an omnidirectional increase in emissivity of functionalized aluminum that results in a hemispherical emissivity near the absolute maximum value of unity in the spectral range of 7.5 to 14 μm. The directional emissivity of a typical optimized FLSP surface is illustrated in Fig. 1a and c. The surface topography is shown in the three-dimensional (3D) laser scanning confocal microscope (LSCM) image and inset scanning electron microscope (SEM) image in Fig. 1b. This sample was processed in an air environment using 35 fs pulses at a 1 kHz repetition rate with a peak fluence of 2.86 J/cm$^2$ and a pulse count of 1600. The microscale surface features typically consist of mounds with heights in the range of 80 to 90 μm. Significant variation in mound diameters are visible in the LSCM image in Fig. 1b. This variation in size is crucial to achieve the broadband high emissivity response. In order to prove that FLSP is highly repeatable to produce near perfect thermal emitters, the optimized surface was reproduced with the same laser processing parameters using two different femtosecond laser systems at three different humidity levels with constant temperature in the lab, in a total of six batches. The hemispherical emissivity ($\varepsilon_h$) value of 0.945 reported in Fig. 1a is the average $\varepsilon_h$ measurement of twelve samples, two per batch. These values were accurately verified by the extensive theoretical analysis presented later in this work. More details about how the hemispherical and directional emissivities are calculated and their definitions are provided in the Supplementary Information section. The standard deviation for the hemispherical emissivity of the twelve samples is also reported in Fig. 1a. Due to the quasi-periodic self-organized nature of the resulting laser processed surface, the exact surface morphology at the microscale varies from one sample to another. However, the macroscale characteristics of the surfaces are uniform and repeatable for a given set of laser processing parameters. The emissivity remained high for a broad spectral range spanning an almost omnidirectional emission angle range, as shown in the measurements presented in Fig. 1c. Note that aluminum oxide has phonon-polariton resonances in the IR wavelength spectrum in the range of interest[59]. The shift of the peak in emissivity from around 11 µm to around 10 µm with increased angle is likely due to a corresponding increase in the oxide thickness based on detection angle.

**Effect of surface structure and oxide thickness on emissivity.** Studies have demonstrated that the background gas used during FLSP has a significant effect on the resulting surface features. For example, processing aluminum in a nitrogen environment has been shown to result in a significant increase in structure height and a reduction in the amount of oxide on the surface compared to structures produced in air[49]. Similarly, the background gas used during processing of silicon has been shown to have a significant effect on the structure shape, underlying chemistry, and the radiative properties[50,60]. The oxide that builds up on the surface structures reported in this paper is likely in the form of oxidized nanoparticles that are created as a result of the laser ablation and deposited on the surface after each laser pulse, similar to the development of aggregated nanoparticle spheres that form using FLSP at low fluence values on aluminum[47,61]. In order to study the effect that the shape of the surface structure has on the emissivity, while maintaining a similar oxide layer thickness between samples, a series of samples were processed in a nitrogen environment with different laser fluences ranging from 0.58 to 4.05 J/cm$^2$. In addition, to study the role of the combination of surface structure and oxides, a series of samples were processed in an air environment for approximately the same range of laser fluences.

LSCM was used to accurately measure the average structure height and surface roughness of each sample (see Table. 1). The reported average height is the average of the maximum height ($R_z$) measured at 10 different areas on the sample[62]. In addition, a comparison between surface oxide layers was accomplished by using a dual-beam system with a scanning electron microscope (SEM) and a focused ion beam (FIB) mill to perform cross sections of the mounds for subsurface analysis of the structures. To prevent damage to the structures during the milling process, a protective platinum layer (PPL), ranging from 2 to 10 µm thick, was deposited first. The cross-sectioned structures were analyzed using energy-dispersive X-ray spectroscopy (EDS) to accurately determine the average thickness of the oxide layer and the mound composition, which is reported in Table 1. Also included in Table 1 are the laser processing parameters, measured surface roughness parameters, and hemispherical emissivity results for each sample. SEM images of cross-sectioned structures for a variety of samples processed in a background gas of nitrogen or air are included in Figs. 2 and 3, respectively. In some of the cross-sectional images, the divisions between layers are difficult to see; in these cases, blue or green lines have been used to better clarify the transitions. In Figs. 2 and 3 different techniques are utilized to image the cross-sectioned structures depending on the sample composition. Imaging with the ion beam highlights elemental contrast. For example, the oxide layer appears very dark as opposed to the aluminum. However, there is significant loss of resolution for imaging with the ion beam versus the electron beam. Use of the electron beam for imaging produces clearer images; however, non-conducting materials (like aluminum oxide) result in a charging effect that washes out the image. Therefore, for samples with a negligible oxide layer, like those illustrated in Fig. 2d-f, SEM images are presented. Whereas for samples with a thick oxide layer, like in Fig. 3d-f, ion beam images are presented instead. Additional SEM images and emissivity data for samples processed with different fluences can be found in the Supplementary Information section 5, in addition to the individual EDS scans for samples presented in Figs. 2 and 3.

All samples processed in nitrogen have a negligible oxide layer thickness of less than 0.5 μm as reported in Table 1. The oxide layer is so thin on these samples that it is not visible in the SEM images in Fig. 2d-f. The EDS surface scan for a sample processed in nitrogen is included in Fig. S8b of the Supplementary Information. EDS surface scans were completed for multiple samples produced in nitrogen and the results were indistinguishable from each other. EDS line scans that were used to identify the different regions outlined in cross-sectional images in Fig. 2 are included in the Supplementary Fig. S9. Because this oxide layer is consistently negligible for the samples processed in nitrogen, it is most likely a result of surface oxidation after the sample has been removed from the nitrogen environment. For the samples processed in nitrogen, as fluence is increased, the roughness and height increase. Furthermore, the thickness of the layer of redeposited aluminum increases with increased fluence. The layer of redeposited aluminum does not contain oxides. From the data in Table 1, as well as the images in Figs. 2 and 3, the hemispherical emissivity increases with increased laser fluence. This also corresponds to an increase in roughness and structure height, until approximately 3 J/cm$^2$. Beyond 3 J/cm$^2$, the roughness and structure height continue to increase, although there is no substantial change in emissivity which is found to plateau or possibly even decrease at higher fluence values.

For samples produced in the air environment, there are some similar trends to the ones produced in a nitrogen environment; in both processing environments, structure roughness and height increase with increased laser fluence as shown in Table 1 and Fig 3. The EDS surface scan for a sample processed in an air environment is included in the Supplementary Fig. S8d. EDS surface scans were completed for the multiple samples produced in air and results were indistinguishable from each other. EDS line scans that were used to identify the different regions outlined in cross-sectional images in Fig. 3 are included in the Supplementary Fig. S10. However, the key difference between the two processing environments can be seen in the redeposited layer thickness. In the air environment, the aluminum nanoparticles that deposit onto the surface after ablation are oxidized and the thickness of the layer of oxidized nanoparticles increases with increased fluence. The importance of the oxidation is illustrated by the dramatically higher hemispherical emissivity values for the samples processed in air rather than nitrogen. For the low fluence values, there are no pits between the mound-like structures (see Fig. 3a) which causes a fairly uniform oxide layer across the sample surface. As the fluence is increased, the size of the pits between each structure increases (see Figs. 3b and c). The oxide layer is thinner in the pits than on the tops of the structures; therefore, the oxide layer is less uniform and thinner on average as the pit size increases, which yields a decrease in the emissivity. The oxide layer thickness on the top of the structures versus the transition into the pits is more clearly depicted in Fig. S7 in the Supplementary Information, which illustrates a broader view of the cross section shown in Fig. 3c and f. This trend is further evidence that the oxide plays a significant role in the high emissivity value of the optimized FLSP surfaces. The crucial role that the oxide layer plays in the emissivity enhancement is also evident by making direct comparison between samples processed in air versus nitrogen. For example, the sample processed in nitrogen at a fluence of 2.86 J/cm$^2$ has an average surface roughness nearly three times greater than the sample produced in air, but the sample processed in air has a higher emissivity. A comparison between the two samples processed in air versus nitrogen at a fluence of 1.14 J/cm$^2$ shows that despite having similar roughness and height, the hemispherical emissivity of the sample processed in air is nearly double compared to the sample processed in nitrogen (see Table 1).

Table 1 Laser processing parameters with corresponding surface roughness parameters and emissivity for FLSP samples processed in either air or nitrogen.

| Peak Fluence (J/cm$^2$) | Pulse Count | Figure | Average Oxide Layer Thickness (μm) | Average Roughness, $R_a$ (μm) | Average Height, $R_z$ (μm) | Hemispherical Emissivity, $\varepsilon_h$ |
|---|---|---|---|---|---|---|
| *Nitrogen* | | | | | | |
| 0.58 | 1865 | 2a | < 0.5 | 6.03+/-0.1 | 57.26+/-0.3 | 0.265+/-0.011 |
| 1.14 | 1865 | - | - | 10.59+/-0.4 | 88.02+/-0.3 | 0.399+/-0.016 |
| 1.85 | 1865 | 2b | < 0.5 | 18.33+/-0.6 | 127.69+/-3.5 | 0.600+/-0.024 |
| 2.23 | 1865 | - | - | 22.56+/-0.7 | 239.83+/-8.6 | 0.815+/-0.033 |
| 2.86 | 1865 | - | - | 35.59+/-1.1 | 317.56+/-14.3 | 0.852+/-0.034 |
| 3.43 | 1865 | - | - | 42.78+/-0.8 | 377.45+/-14.3 | 0.838+/-0.034 |
| 4.05 | 1865 | 2c | < 0.5 | 52.86+/-1.9 | 496.67+/-38.5 | 0.843+/-0.034 |
| *Air* | | | | | | |
| 0.58 | 1865 | 3a | 2.5+/-1.5 | 3.38+/-0.3 | 47.18+/-1.4 | 0.786+/-0.031 |
| 1.14 | 1865 | - | - | 9.97+/-0.8 | 94.01+/-3.8 | 0.865+/-0.035 |
| 2.23 | 1685 | - | - | 10.78+/-0.6 | 120.71+/-6.6 | 0.904+/-0.036 |
| 2.86 | 1865 | 3b | 6.5+/-2.5 | 12.44+/-0.5 | 130.43+/-2.3 | 0.937+/-0.038 |
| 3.43 | 1865 | - | - | 15.81+/-0.9 | 155.04+/-9.2 | 0.936+/-0.037 |
| 4.05 | 1865 | - | - | 22.92+/-0.7 | 179.40+/-5.2 | 0.926+/-0.037 |
| 4.28 | 1865 | 3c | 5.1+/-2.2 | 25.38+/-0.4 | 217.45+/-6.6 | 0.856+/-0.034 |

To examine the effect of the oxide layer thickness on the emissivity more thoroughly, an acid etch technique was used to uniformly remove varying amounts of the surface oxide layer. The etching solution consisted of a mixture of chromic and phosphoric acids, which dissolves aluminum oxide with no significant effect on the underlying metal[63]. The varied parameters for the etch duration and concentration are listed in Table 2, along with the measured average thickness of the oxide layer, surface roughness, and hemispherical emissivity. After etching the samples, mounds of similar size and shape were cross-sectioned. The results on the measured oxide layer thickness are included in Table 2 and Fig. 4. The reported hemispherical emissivity values are the average of four measurements total across two samples for each etching amount, along with the standard deviation. After the acid etching, there is a consistent decrease in the hemispherical emissivity with a corresponding decrease in oxide layer thickness, which is further evidence of the important role the oxide plays in the high emissivity values. There is also an initial decrease in average height with etching; however, the average height remains nearly constant with increased etching beyond the third etch level, while the emissivity continuously decreases along with the decrease in oxide thickness. The decrease in structure height with etching is likely because during FLSP there is preferential redeposition of the oxidized nanoparticle layer on the top of the mounds versus the valleys (or pits). Therefore, during etching more material is removed from the top of the structures than the valleys. There are also only minor changes in the average roughness after the acid etching that do not follow any trends with the changes in the emissivity.

Table 2 Acid etching parameters with corresponding surface roughness parameters and emissivity for FLSP samples all produced using the laser parameters listed at the top of the table.

| *Processing Parameters: open air, fluence = 2.86 J/cm$^2$, pulse count = ~1600* | | | | | |
|---|---|---|---|---|---|
| *% Chromic acid in solution and etch time* | *Figure* | *Average Oxide Layer Thickness (μm)* | *Average Roughness, $R_a$ (μm)* | *Average Height, $R_z$ (μm)* | *Hemispherical Emissivity, $\varepsilon_h$* |
| *no acid etch* | *4a, b* | *8.0+/-2.0* | *9.9+/-0.3* | *89.5+/-5.4* | *0.945+/-0.038* |
| *2% for 20 min* | *4c, d* | *6.0+/-1.2* | *11.1+/-0.2* | *82.8+/-4.7* | *0.864+/-0.035* |
| *2% for 60 min* | *-* | *4.5+/-1.0* | *10.5+/-0.3* | *77.5+/-3.4* | *0.821+/-0.033* |
| *2% for 100 min* | *-* | *2.0+/-0.9* | *10.5+/-0.2* | *77.5+/-3.3* | *0.795+/-0.032* |
| *10% for 60 min* | *4e, f* | *1.3+/-0.8* | *10.5+/-0.2* | *78.2+/-3.2* | *0.783+/-0.031* |

**Theoretical modeling of the laser processed surfaces.** To theoretically demonstrate the effect that the oxide layer and surface morphology have on the emissivity, we perform full-wave electromagnetic simulations utilizing the finite element method software, COMSOL Multiphysics. To this end, we model and compute the thermal emission of a supercell composed of one, two, and three hemispherical mounds with different dimensions and with varied oxide layer thickness. The results of a supercell composed of two mounds are depicted in Fig. 5 and comprehensive results for one, two, and three mounds with varied oxide thickness are included in Supplementary Figs. S12-S14. The supercell mounds are surrounded by periodic boundary conditions at the left and right sides, as shown in Fig. 5b and e. The dispersive properties of aluminum[64] and aluminum oxide[59] are taken from experimental data. Note that aluminum oxide has phonon-polariton resonances at IR frequencies[59], leading to increased losses in this wavelength range and resulting in high emissivity. This resonance is centered around 11 µm and is demonstrated in Fig. 1c. As the angle of emission increases the resonance shifts toward shorter wavelengths because of the changing thickness in the oxide layer.

The radii of the supercell mounds are similar to the mounds shown in the cross sections in Fig. 4a. Considering that the oxide layer is thicker and more homogeneous on the top of mounds compared to the valleys (or pits), the height of the simulated structures is also taken from the Fig. 4a cross sections. Note that the experimentally obtained FLSP surface features are not perfectly periodic and vary in height and shape, but the supercell used was found to be a good approximation to accurately model the presented structures without resorting to the extreme computational burden imposed by modeling random or quasi-periodic elongated surface morphologies. The theoretical results are depicted in Fig. 5. The theoretical simulation results are found to be in near perfect agreement with the experimental results. More specifically, both simulations predict an increase in emissivity over that of a bare flat aluminum surface, which has a negligible hemispherical emissivity of 0.041, as shown in Fig. S2 in the Supplementary Information section. There is also a substantial increase in emissivity for the FLSP surfaces over that predicted for a planar aluminum oxide layer on an aluminum substrate (with results shown in Fig. S11 in the Supplementary Information) which theoretically proves that not only the oxide layer thickness, but also the microscale surface formations, are crucial components that lead to the obtained high emissivity values. Simulations were also preformed to prove that the nanoscale surface features visible in Fig. 1 have no effect on the emissivity in the IR spectrum. The results are illustrated in Supplementary Fig. S15.

The theoretical results of the bare (no oxide) aluminum mounds structure shown in Fig. 5a are comparable to those depicted in Table 2, where the samples with an oxide layer thickness less than 1 μm have a hemispherical emissivity in a comparable range. As the oxide layer thickness is increased on the simulated structure, the hemispherical emissivity rapidly increases. The resulting emissivity for the simulation using an oxide layer with a thickness comparable to the measured value from the cross sections in Figs. 4a and b are included in Fig. 5c, with the supercell structure that was used represented in Fig. 5e. Figure 5d is a large-area 3D schematic of the periodic arrangement of the supercell presented in Fig. 5e. The simulation results with the oxide layer accurately match the experimentally measured values for these surfaces presented in Fig. 1 and Table 2. These theoretical results clearly prove that the cause of the exceptionally high and omnidirectional emissivity is due to the microscale surface formations, in addition to the thick oxide layer formed along the presented FLSP surfaces.

## Discussion

FLSP is an emerging advanced manufacturing technique that can be used to functionalize aluminum surfaces to have broadband omnidirectional hemispherical emissivity close to the absolute maximum value of unity in the spectral range from 7.5 to 14 μm. In addition, the FLSP surfaces have high emissivity even at grazing angles, which is very challenging to achieve with coatings, metamaterials or other perfect emission surfaces. Extensive experimental results along with accurate theoretical modeling demonstrate that there are two key contributing factors to the increase in emissivity; microscale surface roughness and a thick oxide layer that forms when FLSP is applied using the presented processing parameters. Processing in a nitrogen atmosphere results in an increase in surface roughness compared to processing in an air environment using similar processing parameters. However, the thick oxide layer on samples processed in air results in higher emissivity values than samples processed in nitrogen. Therefore, processing in air results in surfaces better optimized for potential applications. The use of an acid etch technique to uniformly decrease the thicknesses of the oxide layer without affecting the underlying structure morphology demonstrates the key role that the oxide layer thickness plays in the high emissivity. The best performing

FLSP surfaces have higher omnidirectional emissivity values than current coatings or metamaterials. They also have additional important benefits that include significantly wider bandwidth and lower fabrication complexity than metamaterials, as well as greater permanency and durability compared to coatings, which is a key property for operation in harsh environments. With the use of industrial high repetition rate ultrashort pulse lasers that are available today, this functionalization technique represents a quick, low-cost, and large-scale fabrication technique without the added weight, hazard of toxicity, and long curing time required in many comparable technologies. The presented FLSP surfaces are ideal for thermal management applications, such as passive radiative cooling, thermophotovoltaics, thermal management of satellites, and other space applications.

## Methods

**Femtosecond laser surface processing.** For laser processing the samples, the experimental setup consisted of a femtosecond laser system, beam delivery and focusing optics, motorized 3D stages, sample environmental chamber, and a computer to control the system (See references for diagram[47,49]). For the samples processed in different background gases, the surface processing was completed in a vacuum chamber attached to the motorized stages with a flow rate of 20-25 scfh of the respective gas at atmospheric pressure. Laser input power was adjusted to account for 8.2% loss from the input window of this chamber. The best performing samples, as well as those used in the acid etching were processed in open air without the vacuum chamber. The femtosecond laser systems used were titanium (Ti):sapphire based amplified systems (a Coherent Inc. Legend Elite Duo and a Coherent Inc. Astrella) generating 35 fs pulses, with a central wavelength of 800 nm, a pulse repetition rate of 1 kHz, and a maximum output pulse energy of 10 mJ and 6 mJ respectively. The laser spot size on the sample was measured by placing a beam profiler with the imaging plane at the same location where the sample is located during processing. The spot size, raster scanning parameters (pitch and velocity), and pulse energy, measured using a thermal pile detector, were used to calculate the peak fluence (the energy per unit area at the peak of the Gaussian) and pulse count. The sample material used was mirror polished aluminum alloy 6061. Before the laser processing, the samples were cleaned in an ultrasonic bath in a 2-step process consisting of a 15-minute ethanol bath followed by a 15-minute deionized water bath. Immediately before each sample was placed in the chamber it was wetted with ethanol and blown dry with nitrogen to remove any surface contamination. After processing, emissivity was evaluated, and the surface structure was characterized by SEM (FEI Quanta 200) and LSCM (Keyence VK-X200K). The LSCM was used to quantify the structure height and average roughness.

**Optimizing emissivity.** In order to systematically study the effects that different processing parameters and background environments have on emissivity, an iterative process was used to find the processing parameters that lead to the maximum hemispherical emissivity. With initial experiments on aluminum that included studies on a wide range of surface structures, it was found that mound-like structures[44] resulted in the highest emissivity values. For these experiments, samples were first produced using a range of laser fluence values from 0.38 to 4.85 J/cm$^2$, with a constant pulse count of around 1900. This process was completed in controlled atmospheres of air and nitrogen as well as an open-air environment. A representative range of resulting surface morphologies and the properties of the surfaces produced in the controlled environments can be seen in Figs. 9 and 10 as part of the Supplementary Information. To achieve the maximized emissivity, the processing parameters were varied slightly around their initial values for the best performing sample. First, pulse count was varied in steps of approximately 10% until reaching a value of about 50% above and below the starting value. Again, the processing parameters from best performing sample was chosen. Next, fluence was varied in steps to reach a value of about 20% above and below the starting fluence to find the best results. Using this process, we found the best results could be produced using a fluence between 2.6 and 2.8 J/cm$^2$ and a pulse count of 1600 to 2000.

**Measuring directional and hemispherical emissivity.** In this work, the hemispherical emissivity is calculated from experimentally measured directional emissivity values using conservation of energy and the Stefan-Boltzmann law (Eqs. 5 and 6 in the Supplementary Information). We utilize a thermal imaging camera (FLIR A655sc) and a sample with a known directional emissivity as the calibrated source. The calibrated source used was a single roll of black polyvinyl chloride electrical tape. The directional and hemispherical emissivities of this tape were quantified using a reflection-based instrument (Surface Optics ET-100). To measure the emissivity, the temperature of the calibrated source and the sample of interest are heated to the same temperature, 50°C. This process helps minimize the contribution of background radiation as well as ensure the samples radiate equal amounts of energy. The heating effect is minimal to the emissivity[65–67]. The thermal imaging camera operates over a spectral range from 7.5 to 14 μm and was used to evaluate the directional emissivity from 0 (normal to the surface) to 85 degrees. The directional emissivity values were used to calculate the hemispherical emissivity using Eq. 7 in the Supplementary Information. Also, further justification of this method is described in the Supplementary Information section. The spectral directional emissivity values presented in Fig. 1c were measured using a reflection-based instrument (Surface Optics SOC-100).

**Acid etching technique.** In order to better understand the role that the oxide layer, introduced by the FLSP process, plays on the resulting emissivity, samples with maximum hemispherical emissivity were etched with an aqueous acid solution consisting of either 20 g/l (2%) or 100g/l (10%) chromic acid and an additional 35ml/l of 85% phosphoric acid solution. During the acid etching the samples were heated to between 82 $^0$C and 99 $^0$C for the specified amount of time. This solution was chosen because it removes aluminum oxide without damaging the underlying metal. Twelve samples from the same batch were used for these studies. Two samples were not etched to use as controls. Six of the samples were etched in a solution of 2% chromic acid in sets of two for different lengths of time at 20, 60, and 100 minutes, respectively. The last two sets of samples were etched in a 10% Chromic acid solution for 60 and 120 minutes, respectively. After etching, the surface morphology and emissivity were re-evaluated. Surface structures were cross-sectioned using FIB milling and then characterized by SEM and EDS (FEI Helios NanoLab 660).

**Theoretical simulations.** The reflectivity spectra of the presented FLSP surfaces were simulated for different incident angle plane waves using the RF module of COMSOL Multiphysics. We utilized periodic boundary conditions surrounding a supercell composed of two different mounds with and without an oxide layer on top. The absorption spectra of the structure for different incident angles were computed, which is equivalent to the emission spectrum for different emission angles at thermal equilibrium due to Kirchhoff's law of thermal radiation[68]. The mounds have similar dimensions to the experimentally produced samples. The aluminum oxide layer thickness that was used is also comparable to the experimental measured values. MATLAB was used to post process the COMSOL raw data and to average the emissivity results for different angle and wavelength values with the goal to calculate the hemispherical emissivity for a variety of different surfaces. Further explanation of the used theoretical method is provided in the Supplementary Information.

## Acknowledgements

This work was supported in part by the National Aeronautics and Space Administration (NASA) Nebraska Space Grant NNX15AI09H. This material is based upon research supported by, or in part by, the U. S. Office of Naval Research under award numbers N00014-19-1-2384 and N00014-20-1-2025. The research was performed in part in the Nebraska Nanoscale Facility: National Nanotechnology Coordinated Infrastructure supported by the National Science Foundation under award no. ECCS: 1542182, and with support from the Nebraska Research Initiative through the Nebraska Center for Materials and Nanoscience and the Nanoengineering Research Core Facility at the University of Nebraska-Lincoln. A special thanks to Scott Hansen at NASA Johnson Space Center for making emissivity measurements using the ET-100 instrument.

## Additional information

**Correspondence** and requests for materials should be addressed to Andrew Reicks, areicks2@unl.edu, Christos Argyropoulos, christos.argyropoulos@unl.edu, and Craig Zuhlke, czuhlke@unl.edu

# Figures

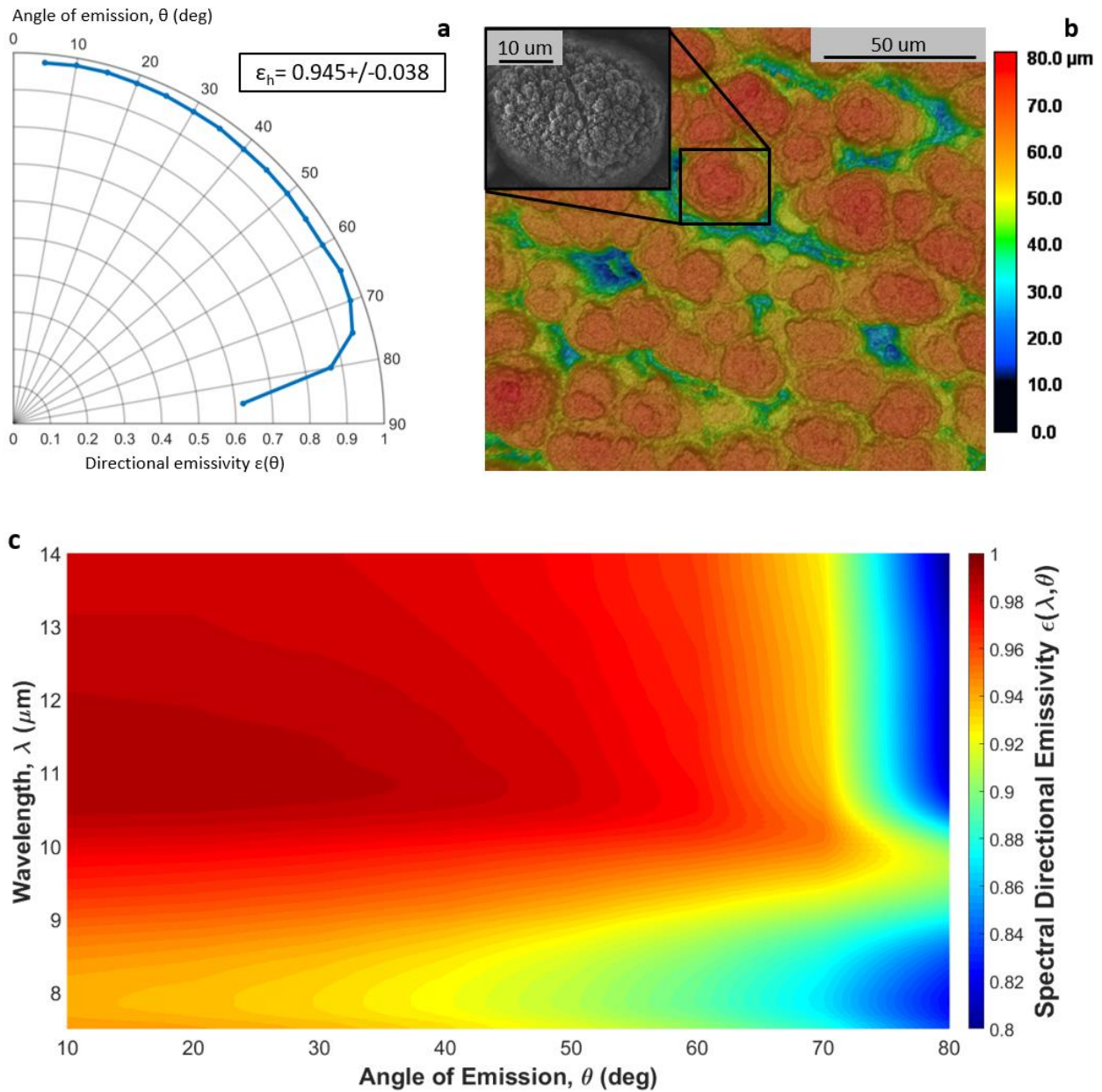

**Fig. 1 Near perfect broadband omnidirectional emissivity response.** (**a**) Directional emissivity as a function of emission angle. Zero degrees corresponds with the detector normal to the surface. The average hemispherical emissivity ($\varepsilon_h$) value is also shown. (**b**) 3D LSCM topographic map of the aluminum laser processed surface with an inset SEM image of a single mound. (**c**) The spectral directional emissivity of the same surface showing that near perfect broadband omnidirectional emissivity response is obtained. Emissivity values were measured every 10 degrees (from 10 to 80) and approximately every 0.1 µm and smoothed using interpolation.

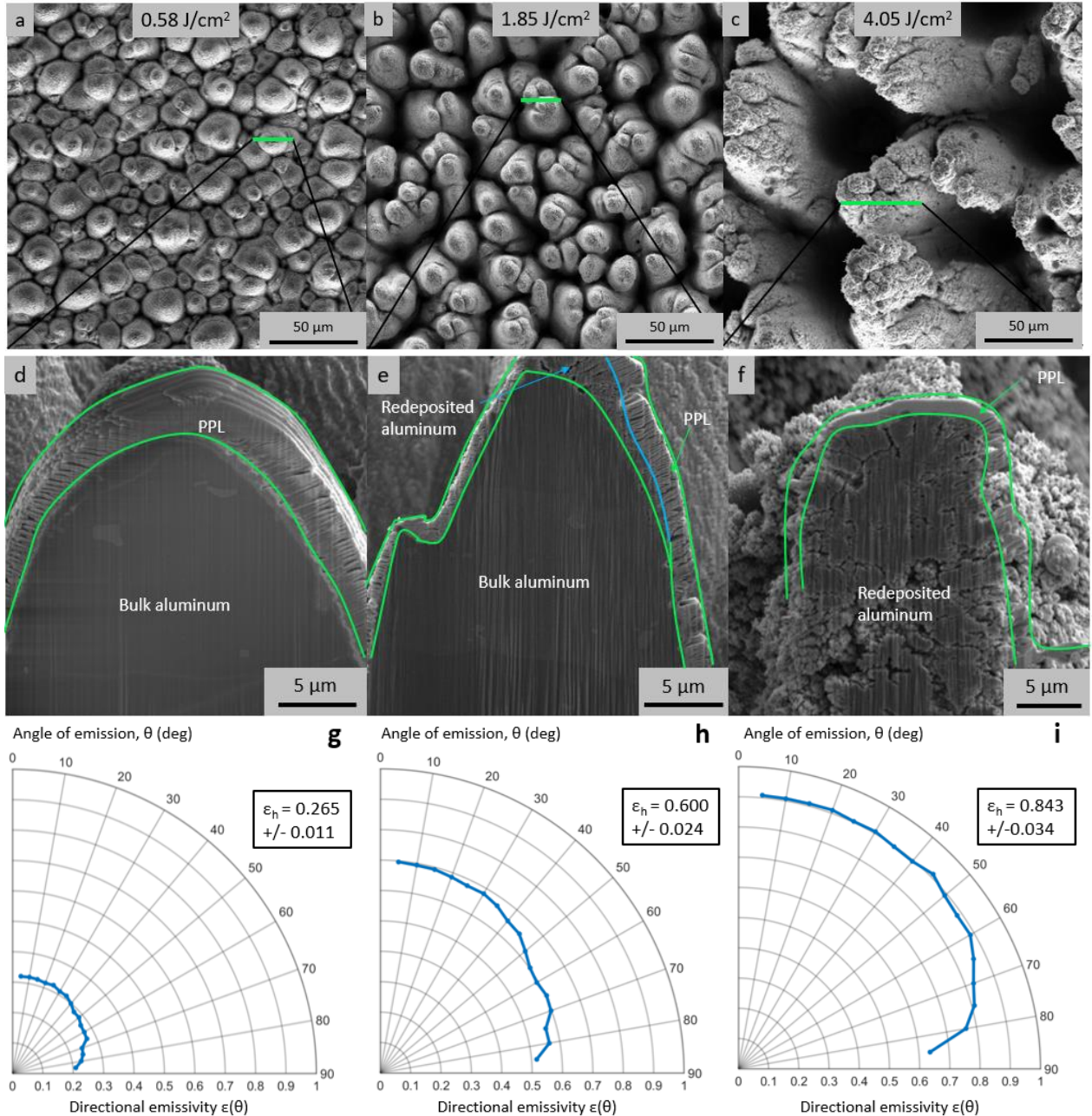

**Fig. 2 Surface and subsurface images, and emissivity of samples produced in nitrogen environment.** (**a-c**) SEM images of samples produced at the fluence specified in the grey box in the top middle of each image for a constant pulse count of 1865. (**d-f**) SEM images of FIB cross-sectioned mounds to show subsurface structure of the corresponding sample in (**a-c**). PPL stands for protective platinum layer that is deposited before the cross-sectioning. (**g-i**) The corresponding directional and hemispherical emissivity of each sample in the same column.

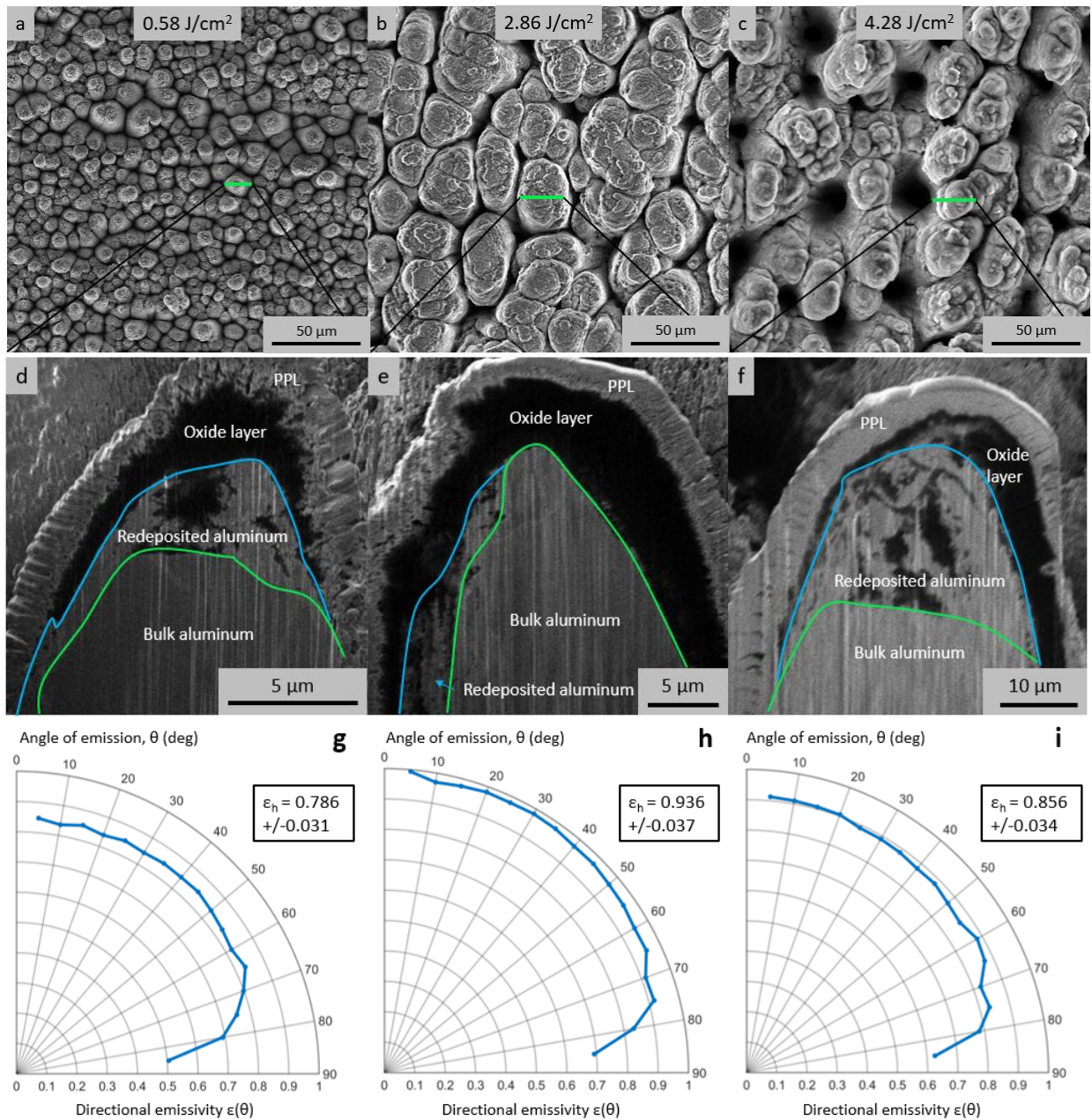

**Fig. 3 Surface and subsurface images, and emissivity of samples produced in air environment.** (**a-c**) SEM images of samples produced at the fluence specified in the grey box in the top middle of each image for a constant pulse count of 1865. (**d-f**) Ion beam images of FIB Cross-Sectioned mounds to show subsurface structure of the corresponding sample in (**a-c**). PPL stands for protective platinum layer that is deposited before the cross-sectioning. (**g-i**) The corresponding directional and hemispherical emissivity of each sample in the same column.

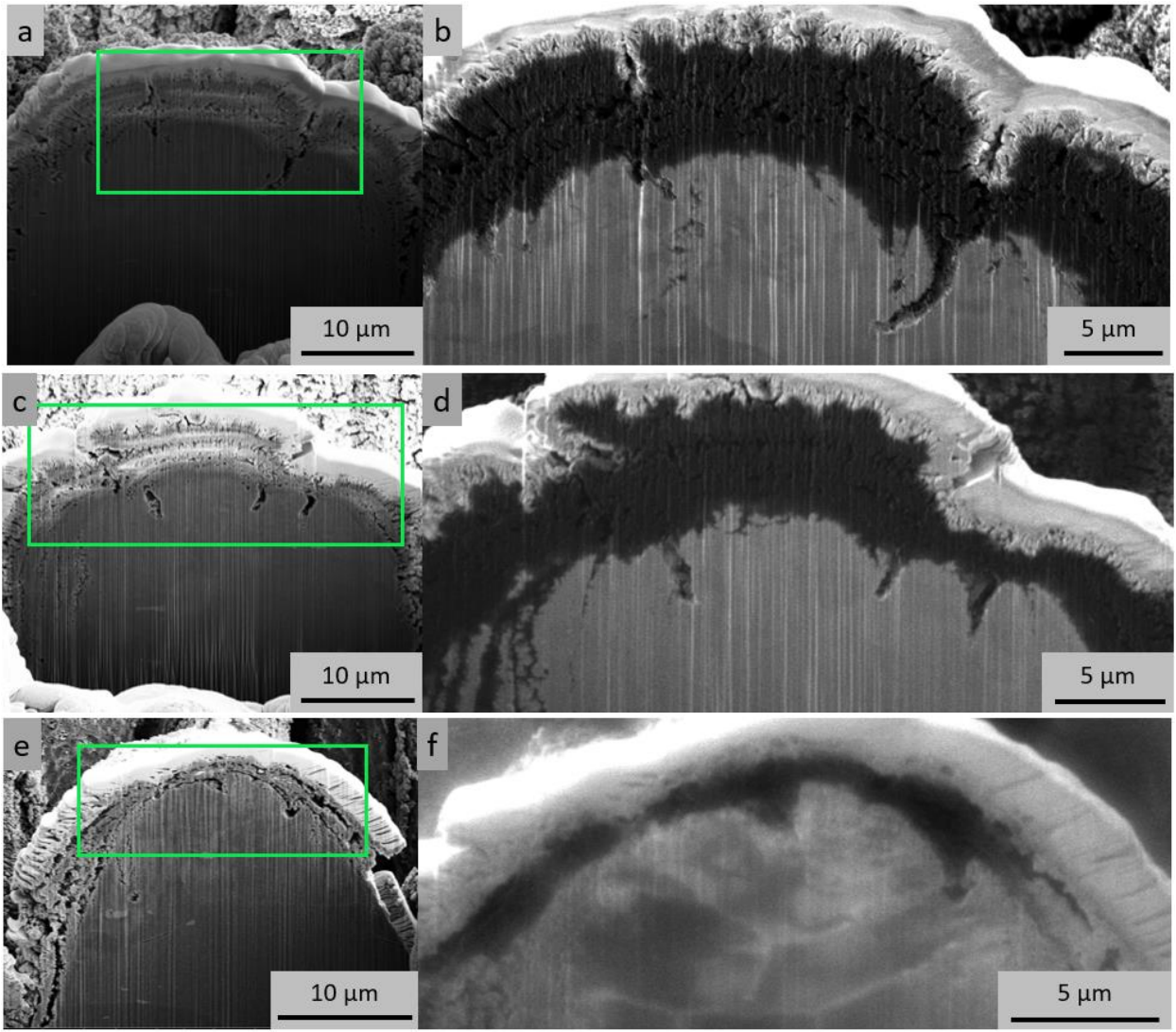

**Fig. 4 Cross-sectional images of three samples processed in open air at the same pulse count and fluence to compare the oxide layer thickness for different amounts of etching**. The cross section in (**a**) and (**b**) was performed directly after laser processing. The remaining samples were acid etched, in a 2% solution for 60 minutes (**c**) and (**d**), and a 10% solution for 60 minutes (**e**) and (**f**). The green boxes in the SEM images of (**a**), (**c**) and (**e**) illustrate the zoomed area for the pictures shown in (**b**), (**d**) and (**f**), respectively. (**b**), (**d**), and (**f**) are images produced using the ion beam as the illumination source, which causes the oxide layer to appear black. The bright layer on top of the oxide is a thin protective platinum layer added before the cross-sectioning.

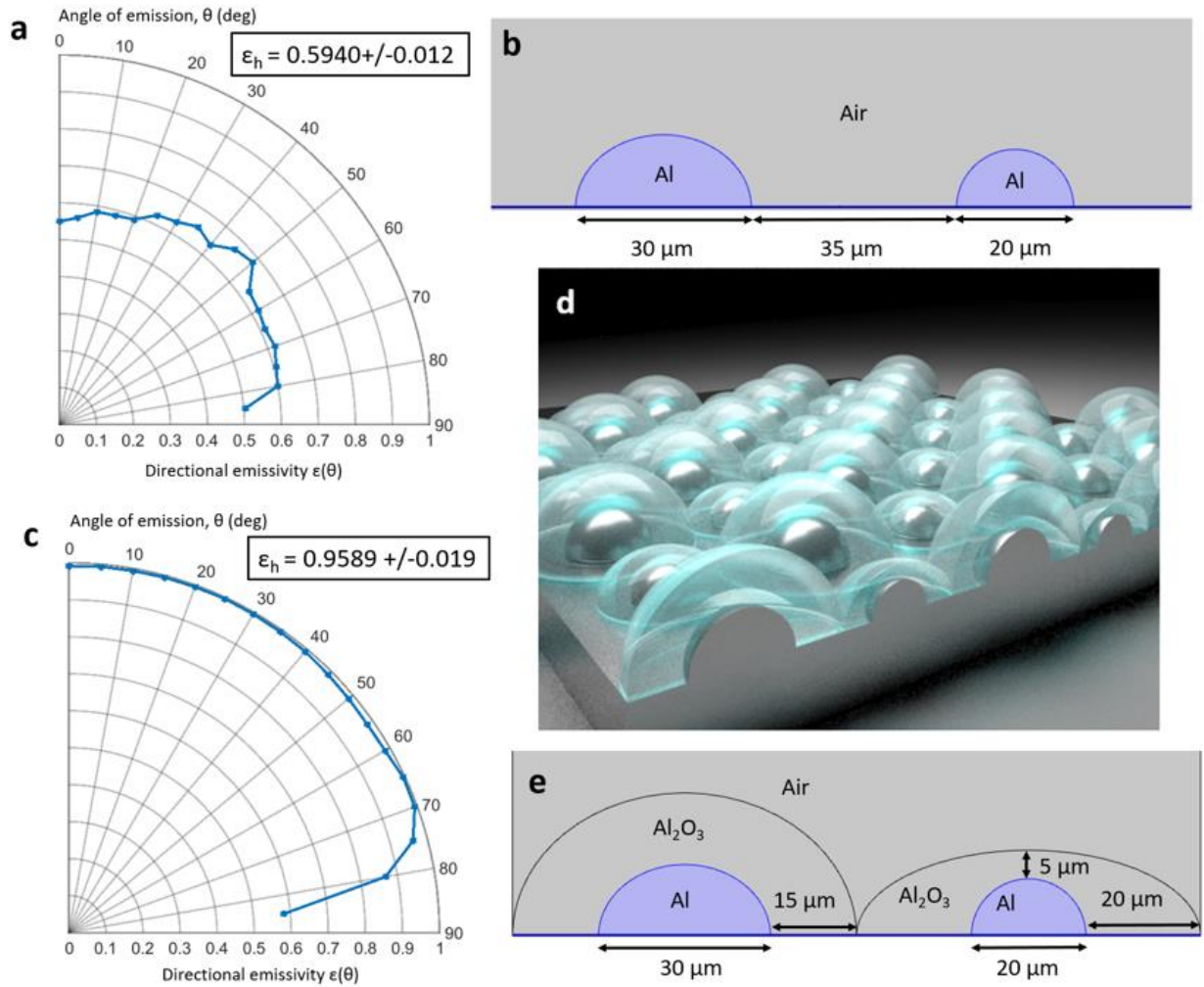

**Fig. 5 Simulations of FLSP surfaces**. (**a**) Simulations of directional and hemispherical emissivity for hemispherical mounds of aluminum with no oxide layer on top. (**c**) Simulations of directional and hemispherical emissivity for hemispherical mounds of aluminum with an oxide layer on top. (**d**) 3D schematic representing a periodic arrangement of the supercell used to calculate the results in (**c**). (**b**) and (**e**) Dimensions of the supercell used in the simulations for the emissivity results shown in (**a**) and (**c**), respectively. The presented simulations accurately agree with the obtained experimental results.

# Supplementary information

## 1. Emissivity of metals and dielectrics

Emissivity is the dimensionless ratio used to describe how efficiently an absorbing surface emits thermal energy, where zero represents a perfect reflector and one represents a perfect emitter. The spectral directional emissivity, $\varepsilon(\lambda, \theta, \varphi, T)$, describes the emissivity of a surface at a particular wavelength, orientation and temperature. The spectral hemispherical emissivity, $\varepsilon(\lambda,T)$, is computed by integrating the spectral directional emissivity over all emission angles at a particular fixed wavelength and temperature. Lastly, the total hemispherical emissivity, $\varepsilon(T)$, is computed by integrating the spectral directional emissivity over all emission directions and in the wavelength range of interest but for a fixed temperature.

Polished metals usually have a low hemispherical emissivity with a higher directional emissivity at low angles relative to the surface normal. When roughness or oxidation is taken into account, the emissivity is usually increased. Generally, the increase of surface roughness leads to an increase in emissivity independent of the wavelength[1,2]. The increase in emissivity caused by roughness is typically illustrated by the optical roughness metric, i.e., the ratio of wavelength divided by the surface roughness. If this ratio is small (less than 0.2), the surface can be described as optically smooth and its properties approach that of an ideal smooth surface with emissivity computed by theory using Maxwell's equations. If this ratio is large, then a geometric optics approach or full-wave electromagnetic simulations must be utilized to take into account the surface morphology[3]. Emissivity is also a function of the surface temperature, wavelength, and observation angle. The effect of these properties can vary greatly depending on the material.

In metals the effect of temperature on emissivity is primarily dependent on the temperature dependent resistivity of the material. The Hagen-Rubens relation shows that for most materials the emissivity is proportional to the square root of resistivity, for sufficiently short wavelengths[2]. Specifically, for aluminum, experimental data has shown that over the temperature range of 0 ºC to 400 ºC the resistivity can be approximated by a linear equation. The resulting effect on emissivity is weak and causes a variance of approximately 0.006 to 0.008 per hundred degrees Celsius[4–6]. Since this value is so small over such a wide temperature range, the effect of temperature on emissivity is ignored here. For wavelengths longer than 1 μm the directional emissivity of metals tends to increase at large angles, leading to a "flat top" profile (an example of this effect is demonstrated in Fig. S2).

For dielectric materials like metal oxides, temperature typically has even less effect on emissivity than for metals. The spectral properties of dielectrics change very slowly with temperature since the refractive index is not a strong function of temperature. For dielectrics, the most significant effect of temperature is related to measuring their thermal radiation power because the wavelength shift in the blackbody radiation distribution needs to be considered[2]. The spectral range considered for this paper corresponds to the atmospheric window from 7.5 to 14 μm or peak blackbody radiation from -66 °C to 110 °C. Unlike metals, in dielectrics the directional emissivity tends to decrease at large angles. However, this can vary greatly with surface roughness.

The FLSP surfaces are a combination of dielectric and metal materials and both must be considered for understanding the increase in emissivity for the processed surfaces. The base aluminum is pure metal and the surface oxide is a dielectric. The thickness of the oxide layer varies greatly depending on the processing parameters.

## 2. Theoretical evaluation of total hemispherical emissivity of a surface

In order to provide the appropriate theoretical background on the different emissivity notations, a mathematical representation is included here. A diagram showing the measurement setup is illustrated in Fig. S1. The spectral directional emissivity, $\varepsilon(\theta, \varphi, \lambda, T)$, of an opaque material is obtained in accordance with Kirchhoff's Law of thermal equilibrium shown in Eq. 1:[2]

$$\varepsilon(\theta, \varphi, \lambda, T) = \alpha(\theta, \varphi, \lambda, T) = 1 - \rho(\theta, \varphi, \lambda, T), \qquad (1)$$

where $\alpha$ and $\rho$ are the spectral directional absorptivity and reflectivity, respectively. Most methods for calculating the emissivity of a surface are derived from these equations by computing the reflectance and assuming no dependence on the solid angle φ, which means the spectral directional emissivity is assumed to be independent of the sample rotation. The spectral hemispherical emissivity, ε(λ, T), can be calculated from the spectral directional emissivity by using the following formula:[2]

$$\varepsilon(\lambda, T) = 2 \int_0^{\pi/2} \varepsilon(\theta, \lambda, T) \sin\theta \cos\theta \, d\theta. \qquad (2)$$

The total hemispherical emissivity, ε(T), is obtained via integration over the Planck distribution (P):[2]

$$\varepsilon(T) = \frac{\int_0^\infty \varepsilon(\lambda, T) P(\lambda, T) d\lambda}{\int_0^\infty P(\lambda, T) d\lambda}. \qquad (3)$$

P is given by Eq. 4:

$$P(\lambda, T) = \frac{8\pi hc}{\lambda^5 e^{\frac{hc}{\lambda kT}} - 1}, \qquad (4)$$

where h is Planck's constant, k is the Boltzmann constant, c is the speed of light in vacuum, λ is the wavelength, and T is the temperature. It is important to note that the integral in Eq. 3 is evaluated from 0 to infinity. However, experimentally it is not possible to make measurements that cover all wavelengths and, therefore, a finite wavelength range must be used. In Eq. 3, the limits of integration correspond to the range of measurement (7.5- 14 μm).

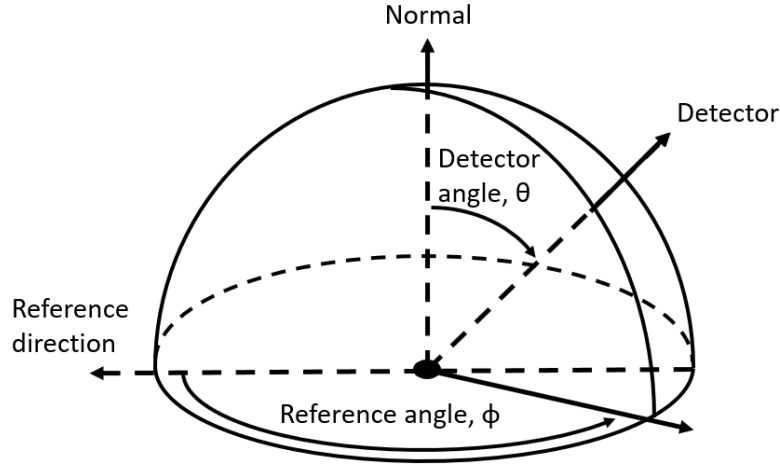

**Fig. S1 Depiction of emissivity measurement setup**. In our case, the origin represents the sample and the detector is the thermal imaging camera. Samples were tested to verify no dependence of emissivity on the reference angle.

## 3. Measuring directional and hemispherical emissivity

In this work, the hemispherical emissivity is calculated from the experimentally measured directional emissivity by using conservation of energy and the Stefan-Boltzmann law (Eqs. 5 and 6). Utilizing the measured temperature of the calibrated source and its emissivity, the temperature and thus the energy of the detector can be found. From here, the directional emissivity of the unknown sample can be calculated as a ratio of the temperature of the sample to that of the detector minus some small background contribution[7] (see Eq. 5). The energies (E) of the detector, sample, and background are calculated from their temperatures by using the Stefan-Boltzmann law (see Eq. 6):

$$E_{detector} = \varepsilon E_{Sample} + (1-\varepsilon)E_{background}, \qquad (5)$$

$$E = \sigma T^4. \qquad (6)$$

Testing was performed to show that the sample reference direction ($\varphi$) had no effect on the emissivity (see Fig. S1 for a depiction of the emissivity measurement setup). The hemispherical emissivity, $\varepsilon_h$, is calculated by using Eq. 7. Note that the only difference between Eqs. 7 and 2, the equation for calculating the spectral hemispherical emissivity, is the lack of a spectral dependence in Eq. 7.

$$\varepsilon_h = \varepsilon(T) = 2 \int_0^{\pi/2} \varepsilon(\theta, T) \sin\theta \cos\theta \, d\theta \qquad (7)$$

However, experimentally we make measurements at discrete angles (not as a continuous function) and the integral described in Eq. 7 must be approximated. For the approximation we employ two methods, the rectangular[8] and trapezoidal integration approximation, and use the average between them. The rectangular approximation tends to overestimate the area of a concave down curve and underestimate concave up, whereas, the trapezoidal approximation has the opposite effect. The difference in these two estimations is used to find the geometric uncertainty in the hemispherical emissivity.

## 4. Verifying our method for measuring emissivity

Two approaches were used to verify the validity of the presented technique for measuring emissivity. First, the hemispherical emissivity of three pieces of mirror- polished aluminum 6061 with an average surface roughness of less than 0.5 µm was measured three times and averaged. The resulting experimental values are reported in Fig. S2. They are typical of mirror polished aluminum and found to be in good agreement with experimental results from the literature ranging between 0.04 to 0.09[2,3,5,6]. Simulation results for the emissivity of a flat aluminum surface as well as analytical values calculated using the equations in Ref.[2] are also included in Fig. S2. The 23% difference in hemispherical emissivity between the experimental and theoretical values is likely a result of the native oxide layer formed on all aluminum surfaces, as well as the surface roughness. Both these effects are not included in this theoretical modeling.

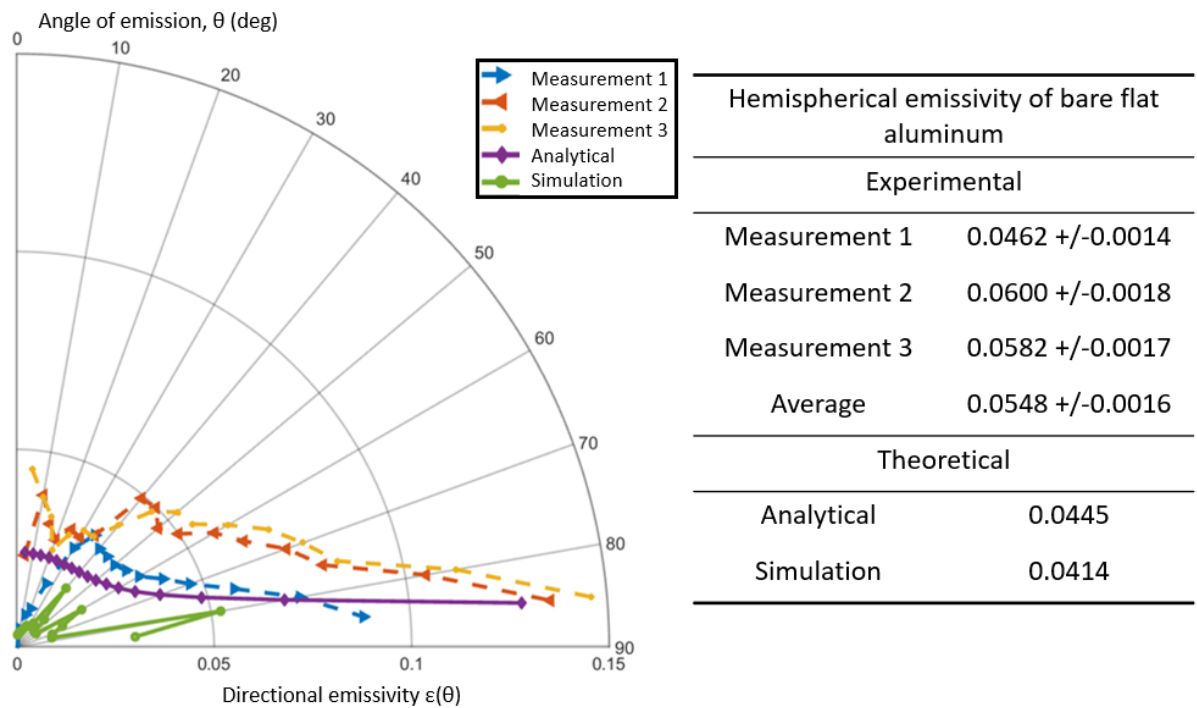

**Fig. S2 Experimental and theoretical values for hemispherical emissivity of bare flat aluminum**. Left: Experimental directional emissivity of bare aluminum measured using the thermal camera method as compared to theoretical values. The analytical and simulated data does not account for the surface roughness of the bare aluminum. Right: Experimental hemispherical emissivity compared to the analytical and simulated values. The analytical value was calculated using the equations in Ref.[2].

The calibrated source used was a single roll of black polyvinyl chloride electrical tape. The reflection-based instrument (Surface Optics ET-100) was used to measure its hemispherical emissivity and directional emissivity at 20 and 60 degrees. These three values were used to fit a curve made by taking the average of 40 measurements of the tape's temperature in increments of five-degree angles to compute the absolute emissivity values. Because the sample is in thermal equilibrium, the apparent change in temperature measured by the thermal camera for different angles is actually a change in emissivity (see Eqs. 5 and 6). Fitting the temperature measurements from the thermal camera to the emissivity measurements with the ET-100 results in the directional emissivity curve for the tape illustrated in Fig. S3.

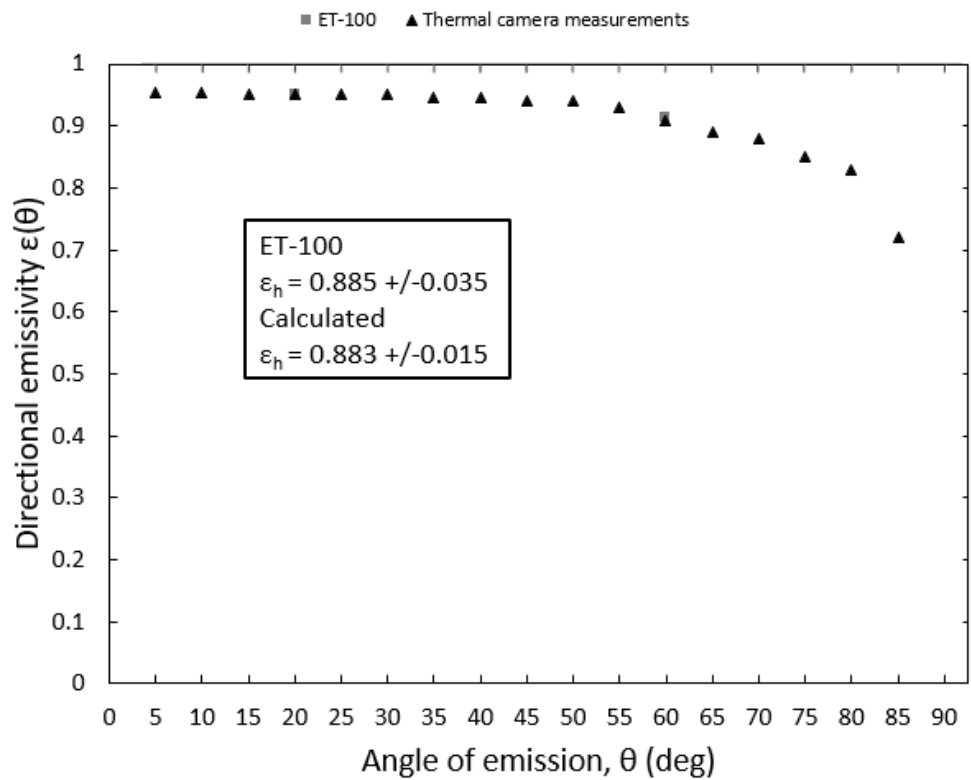

**Fig. S3 Directional and hemispherical emissivity of the black polyvinyl chloride electrical tape used as the calibrated source.**

In addition, the emissivity of the best performing sample (described in the Results section (see Fig. 1)) was measured using a reflection-based instrument (Surface Optics SOC-100), which provides the reflection coefficient as a function of wavelength. The results were compared with our thermal imaging camera (TIC) testing method and are illustrated in Fig. S4. The difference between the measured hemispherical emissivity for the two techniques is negligible and less than 0.5%.

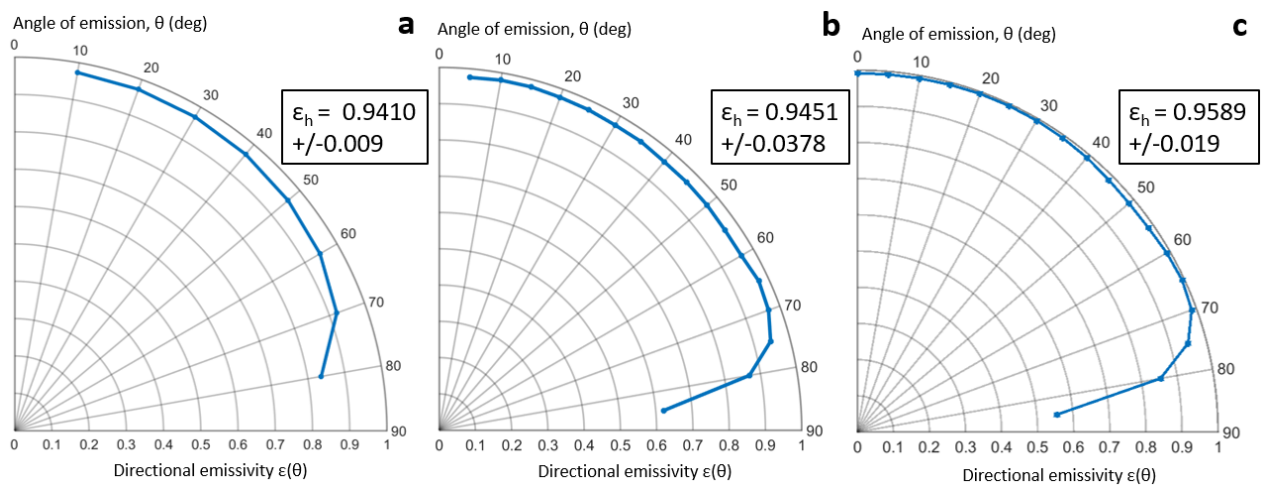

**Fig. S4 Directional and hemispherical emissivity of the best performing surface shown in Fig. 1 in the main paper.** Measurements made using the (**a**) reflection method, (**b**) TIC method, and (**c**) simulations. All three methods are described in more detail in the Supplementary Information section and they are in agreement within the uncertainty values.

## 5. Extended measurement data

As noted in the Results section, directional and hemispherical emissivity measurements and SEM images for additional surfaces processed with varied fluence and processed either in nitrogen or in air are included in Figs. S5 and S6, respectively. The cross section in Fig. S7 is included as an example to demonstrate how, at relatively high fluence values, the oxidized nanoparticles redeposit in a non-uniform manner resulting in the emissivity drop that occurs with increased fluence beyond 3 J/cm$^2$. Energy Dispersive X-Ray Spectroscopy (EDS) surface area scans for both atmospheres are included in Fig S8 to demonstrate the dramatic difference in surface composition between the two environments. Based on the EDS surface scans, processing in nitrogen appears to result in reduced oxygen content compared to processing in air. Figures S9 and S10 include one EDS line scan for each of the cross sections presented in Figs. 2 and 3 in the main paper. This line scan data is included to demonstrate the chemical analysis performed to determine the composition of the layers formed during processing.

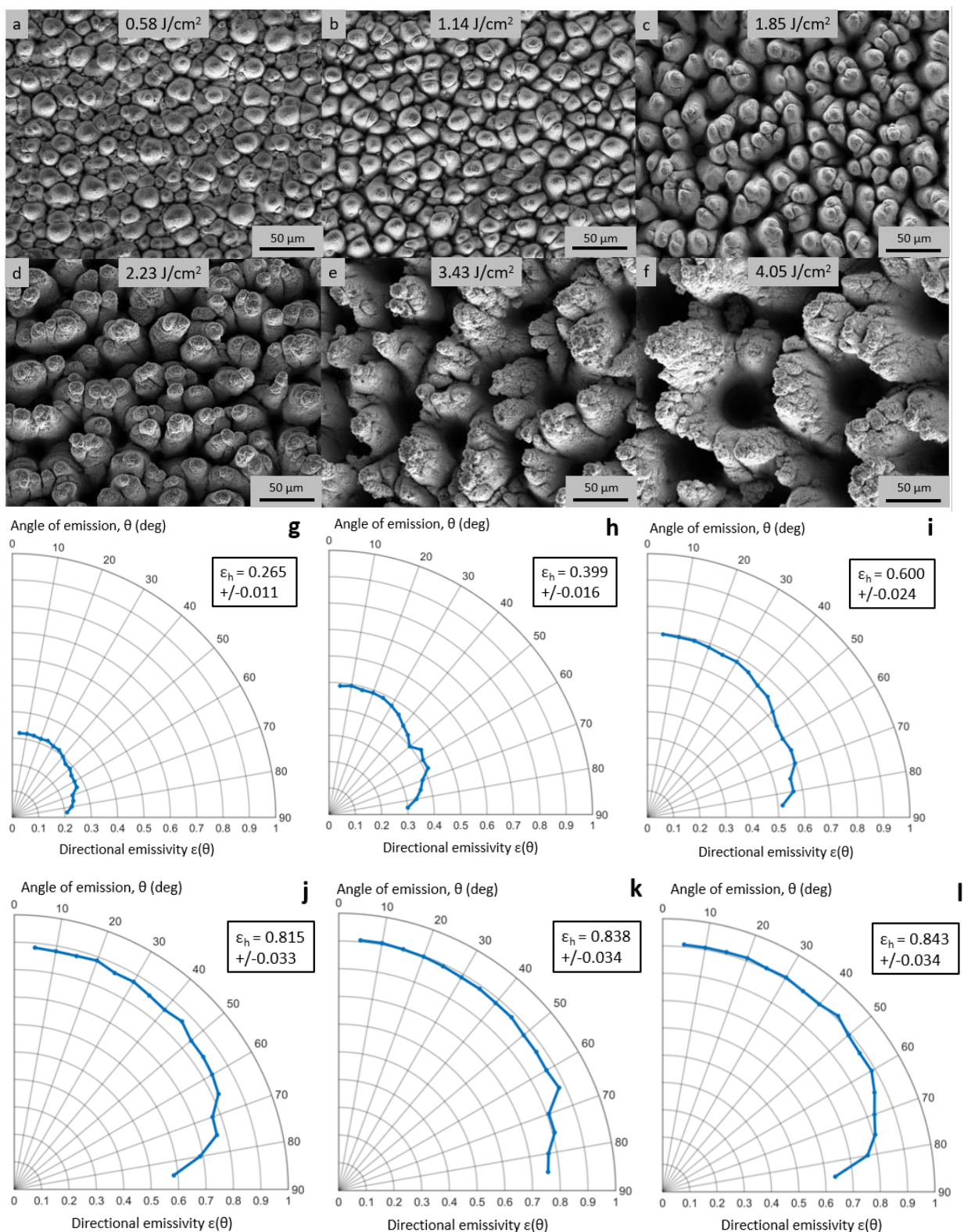

**Fig. S5 SEM images, as well as directional and hemispherical emissivity of surfaces processed in nitrogen**. (**a-f**) SEM images of samples produced at different fluences for a constant pulse count of 1865. (**g-l**) The directional and hemispherical emissivity of the samples. The SEM image in (**a**) corresponds to the emissivity plot in (**g**), (**b**) corresponds to (**h**), and so on.

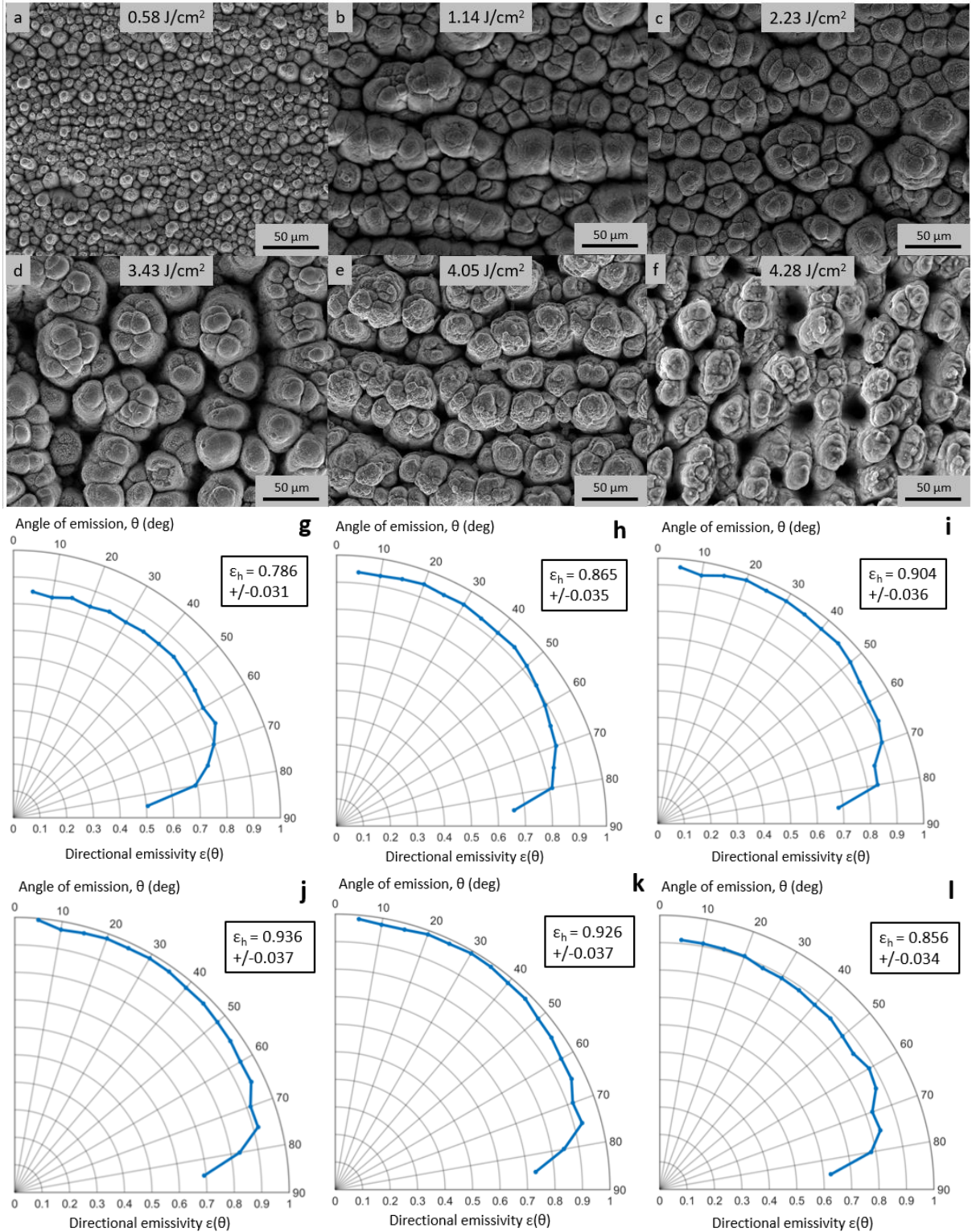

**Fig. S6 SEM images, as well as directional and hemispherical emissivity of surfaces processed in air.** (**a-f**) SEM images of samples produced in air at different fluences for a constant pulse count of 1865. (**g-l**) The directional and hemispherical emissivity of the samples. The SEM image in (**a**) corresponds to the emissivity in (**g**), (**b**) corresponds to (**h**), and so on.

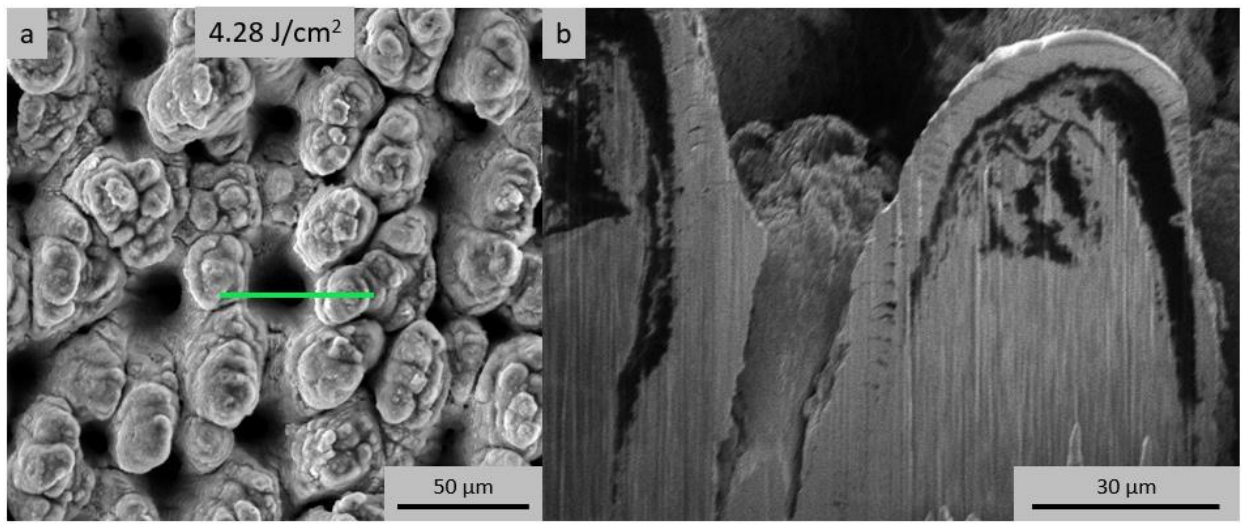

**Fig. S7 Surface and subsurface images for a sample produced in an air environment to show the decrease in thickness of the oxide layer moving down into the pit.** (**a**) Surface SEM image of a sample produced at the fluence specified in the grey box in the top middle of the image for a constant pulse count of 1865. (**b**) Ion beam image of the area through a pit between two mounds (green line in (**a**)) that was cross sectioned with the FIB mill to show the subsurface structure.

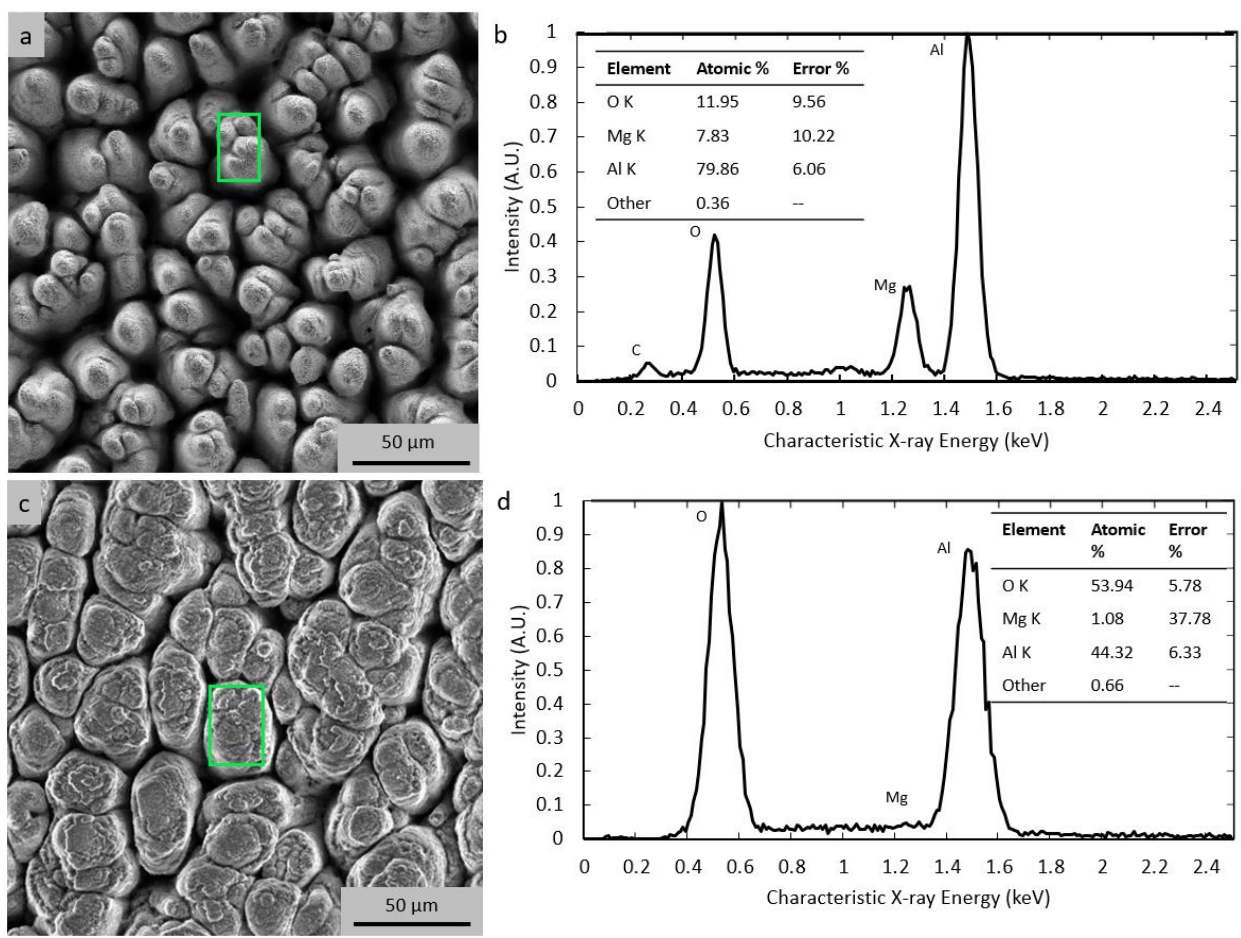

**Fig. S8 EDS surface scans for samples produced in each environment.** The area scan EDS data on the right was collected over the area within the green boxes in the SEM images on the left. (**a, b**) Results for a sample processed in nitrogen at a fluence of 1.85 J/cm$^2$ and a pulse count of 1865 with an oxide layer less than 0.5 µm thick. (**c, d**) Results for a sample processed in air at a fluence of 2.86 J/cm$^2$ and a pulse count of 1865 with an oxide layer of 6.5+/-2.5 µm thick. The lower magnesium content for the sample processed in air (**c, d**) is likely because signal is only being collected from the thick oxide layer. For the samples processed in nitrogen (**a, b**) the oxide is less than 500 nm thick, so some signal is collected from the bulk Al 6061 material, which contains Mg as an alloying element.. The EDS results are indistinguishable between samples produced in the same environment.

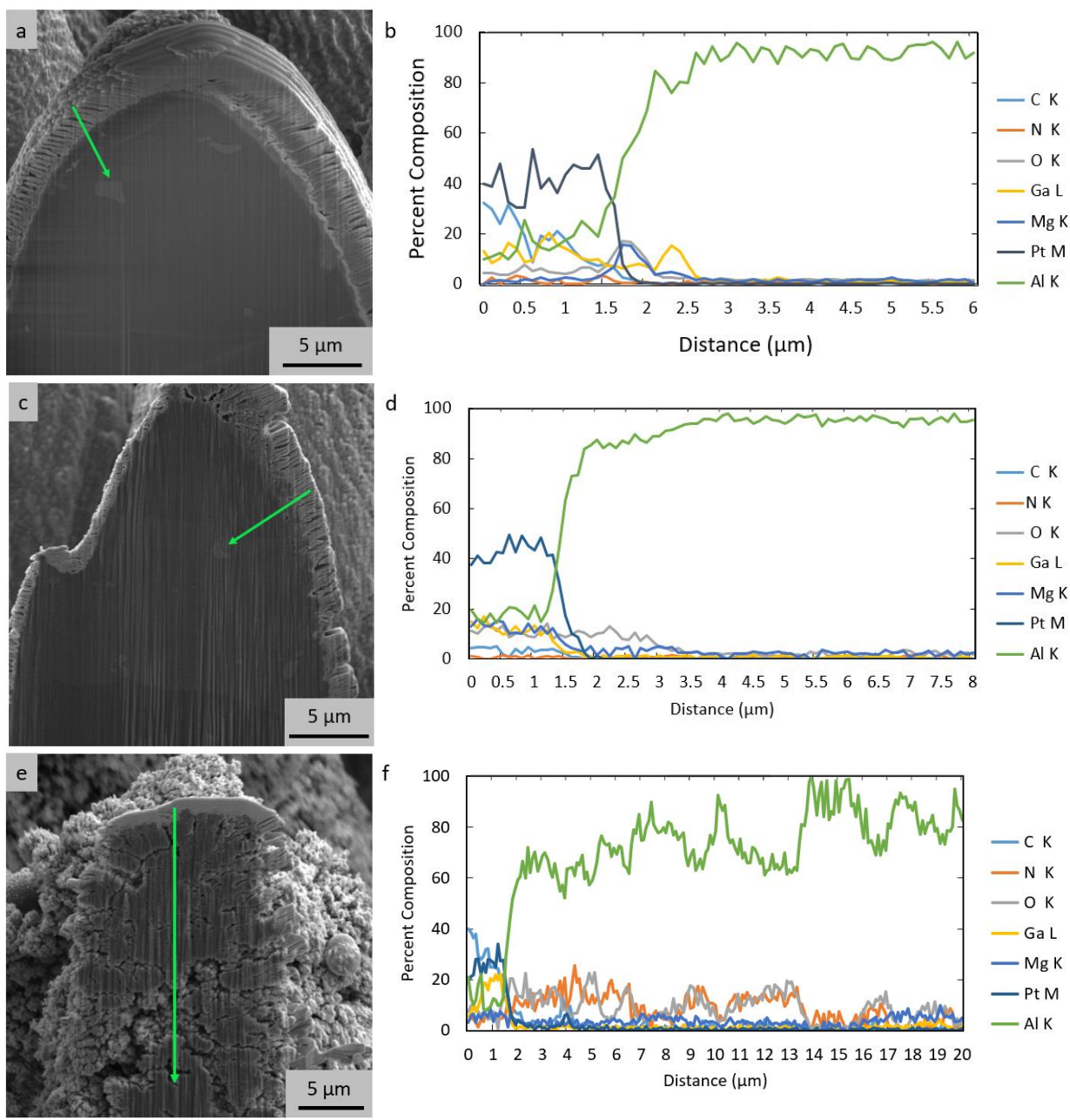

**Fig. S9 EDS line scans for samples produced in a nitrogen environment.** The cross-sectional SEM images on the left correspond with Fig. 2 in the main paper. The green arrows in the SEM images on the left indicate the scan path that corresponds with the EDS line scan data on the right. The samples were processed with a fluence of 0.58 J/cm$^2$ (**a-b**), 1.85 J/cm$^2$ (**c-d**), and 4.05 J/cm$^2$ (**e-f**), and a pulse count of 1865. The alternating bands of relatively high aluminum versus oxygen seen in (**e-f**) are caused by the porosity.

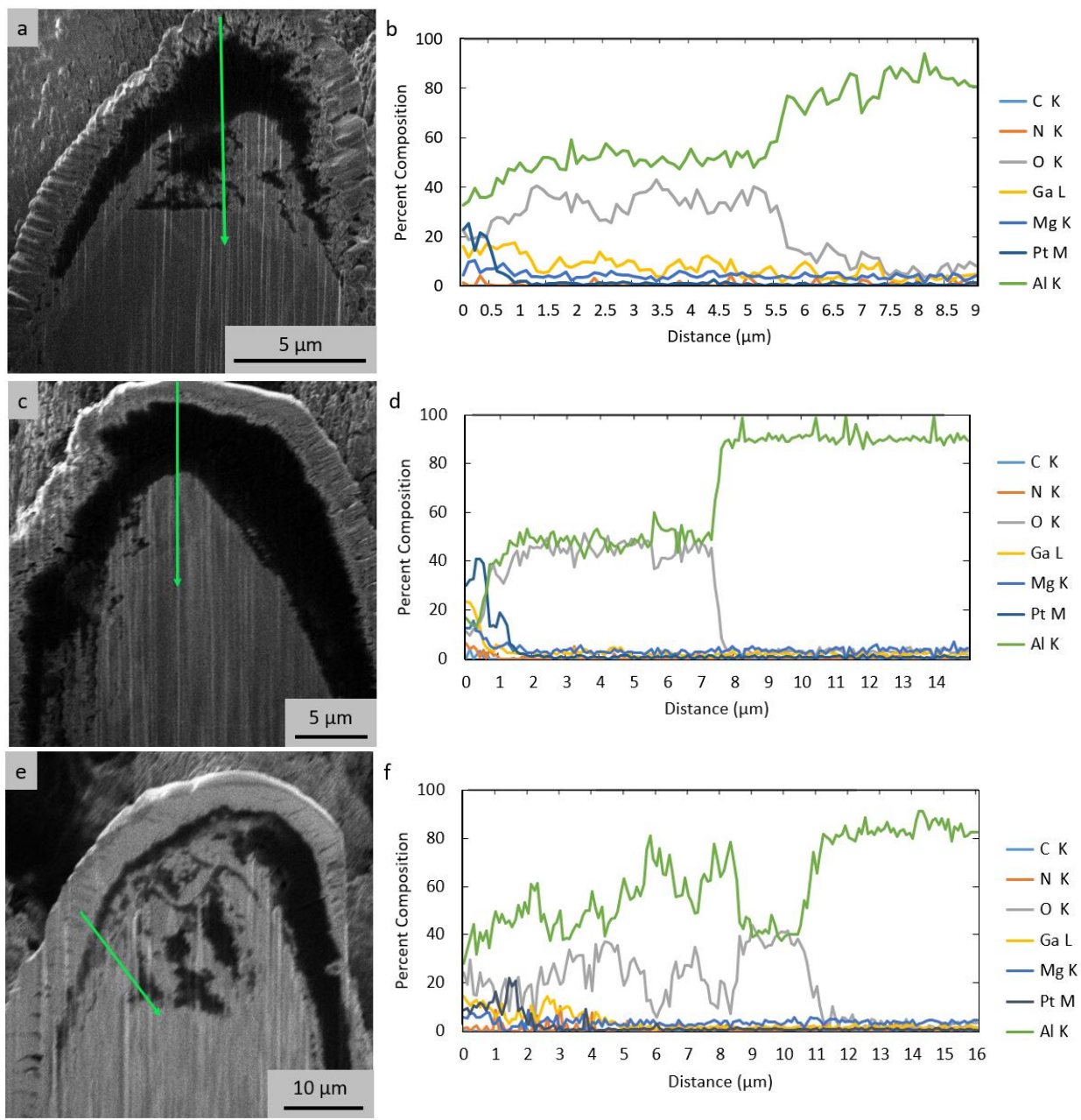

**Fig. S10 EDS line scans for samples produced in an air environment.** The cross-sectional SEM images on the left correspond with Fig. 3 in the main paper. The green arrows in the SEM images on the left indicate the scan path that corresponds with the EDS line scan data on the right. The samples were processed with a fluence of 0.58 J/cm$^2$ (**a-b**), 2.86 J/cm$^2$ (**c-d**), and 4.28 J/cm$^2$ (**e-f**), and a pulse count of 1865.

## 6. Theoretical simulations of a flat aluminum surface

Simulations of an ideal polished flat aluminum surface can be found in Fig. S2 and agree with experimental and analytical results found in the relevant literature[2–6,9–13]. The emissivity of aluminum with an oxide layer on top was also simulated to further verify the theoretical model. Polished aluminum can be anodized to grow a thick oxide layer on its surface. Experimental data shows that the hemispherical emissivity of the aluminum/aluminum-oxide system increases rapidly until the oxide thickness of about 15 μm, where it levels out, asymptotically approaching ~0.85 for larger oxide thicknesses[11–13]. Simulations, illustrated in Fig. S11, were performed for 5, 15, and 20 μm oxide layer thickness and agree with the experimental data reported in the literature.

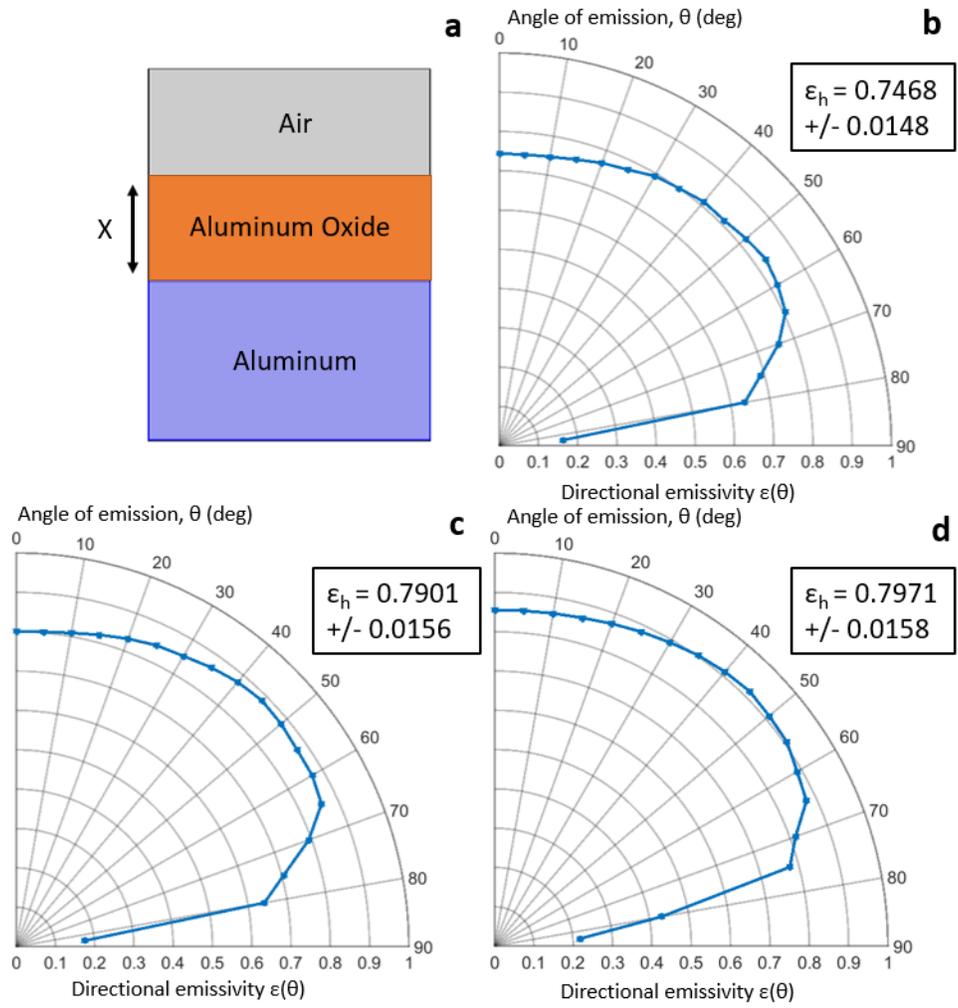

**Fig. S11 Theoretical simulations of hemispherical and directional emissivity for a flat thick aluminum film with varying thickness of top oxide**. (**a**) The schematic of the system used in the simulations. Simulation results for an oxide layer thicknesses of (**b**) 5 μm, (**c**) 15 μm, and (**d**) 20 μm.

## 7. Theoretical study of different structural morphologies effects on the emissivity

In this section we present a comprehensive theoretical study to provide additional physical insights on the influence mechanism of the structure on the emissivity properties of different FLSP surfaces. Towards this goal, we perform full-wave rigorous electromagnetic simulations (similar to Fig. 5 in the main paper) to accurately demonstrate the effects of diverse structural morphologies on the emissivity of different FLSP surfaces. More specifically, we study the emissivity dependence to both the thickness of aluminum oxide, as well as the number of mounds with different sizes produced via the FLSP process. The dimensions of the simulated FLSP structures follow the sizes of the mounds in the experimental cross-sectional images shown in the main paper. The emissivity is again computed in the mid-IR range and its average value is plotted as a function of different emission angles, similar to Fig. 5 and the other experimental results presented in the main paper.

We start our theoretical studies by simulating a single aluminum mound surrounded by periodic boundary conditions at the left and right sides. This consists a simplified unit cell of the usually complex FLSP surface morphology (Fig. 1b in main paper). The computed directional emissivity results with increasing aluminum oxide thickness are shown in Fig. S12. In all these simulations, we keep the diameter of the hemispherical aluminum mound to the fixed value of 30 μm. Interestingly, the emissivity is very limited and low in the case for a bare aluminum mound with no oxide formed on top, as shown in Fig. S12a. As the thickness of the aluminum oxide layer increases, the directional emission is also enhanced, which is evident by the remaining plots in Fig. S12. From these results, it can be concluded that for oxide layer thicknesses larger than $5\mu m$ the emissivity is almost perfect, at least until emission angles of 70°. This response is consistent with the experimental results shown in the main paper.

Next, we compute the emissivity of a supercell, similar to Fig. 5e in the main paper, but now composed of two and three hemispherical mounds with different dimensions and with varied oxide layer thicknesses. The relevant directional emissivity results along with the simulation schematics are shown in Figs. S13. and S14, respectively. Again, we keep the diameters of the hemispherical aluminum mounds fixed in all simulation models and equal to $a_1 = 30$ μm, $a_2 = 20$ μm, and $a_3 = 10$ μm. To ensure the same rate of oxide growth on top of each aluminum mound, the thickness of each oxide layer is changed proportionally to the oxide thickness added on the first mound, $d_1$, by using the relations: $d_2 = a_2 \cdot d_1/a_1$ and $d_3 = a_3 \cdot d_1/a_1$. Directional emissivity enhancement is obtained by increasing the thickness of aluminum oxide on top of the different morphologies of aluminum mounds, as is evident by Figs. S13 and S14. However, in these supercell cases, the emissivity is almost perfect for emission angles even higher than 70º, on the contrary to the single mound design shown in Fig. S12. This is the main reason that a supercell was used in the simulations of the main paper (Fig. 5) instead of a single mound. Hence, it can be concluded that the supercell geometry (either two or three mounds) follows more accurately the morphology and dimensions of the experimentally obtained quasiperiodic FLSP surfaces.

Finally, it is worthwhile to mention that perfect emissivity from the FLSP surfaces is only due to the micrometer-scale geometrical features and the oxide layer and is not related to the nanoscale features. We checked this issue by modelling the single mound geometry at the nanometer scale with nanoscale dimensions ($a_1 = 30$ nm and $d_1 = 15$ nm) and concluded that the emissivity is very low, as shown in Fig. S15. The nanometer-scale geometrical features of the FLSP surfaces do not play a role in the computed mid-IR emissivity since the wavelength range of this mid-IR response is in the tens of micrometer scale (7.5 to 14 μm), which is tens to thousands of times larger than the nanometer scale morphological features.

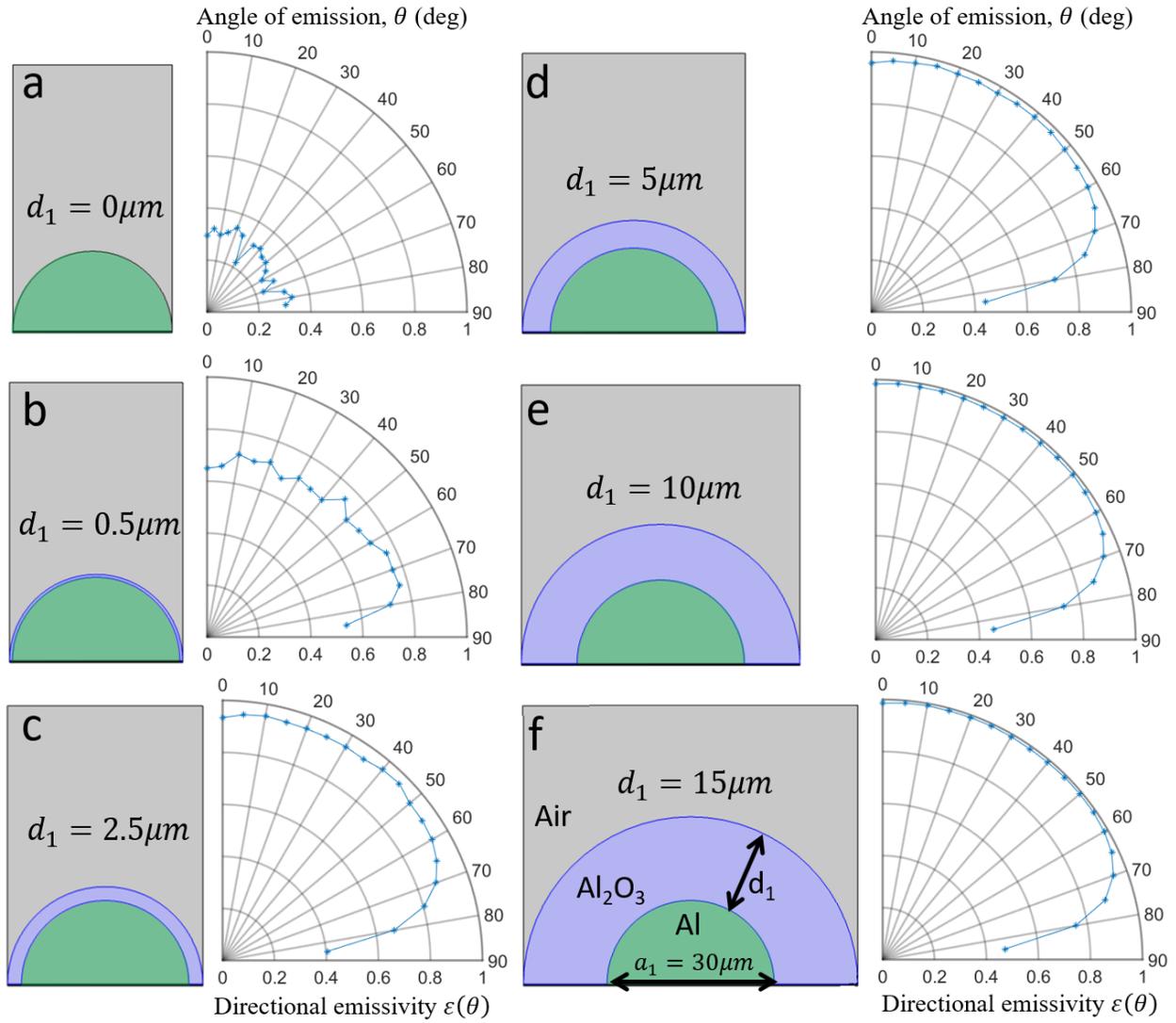

**Fig. S12 The emissivity of a single periodic aluminum mound with varying aluminum oxide thicknesses**. (**a-f**) Left panels: schematics of a single periodic aluminum mound with varying aluminum oxide thicknesses. Right panels: computed directional emissivity values associated to different thicknesses of the aluminum oxide layer ranging from 0 µm in (**a**) (bare aluminum mound) to 15 µm in (**f**). The aluminum mound has a fixed diameter of 30 µm.

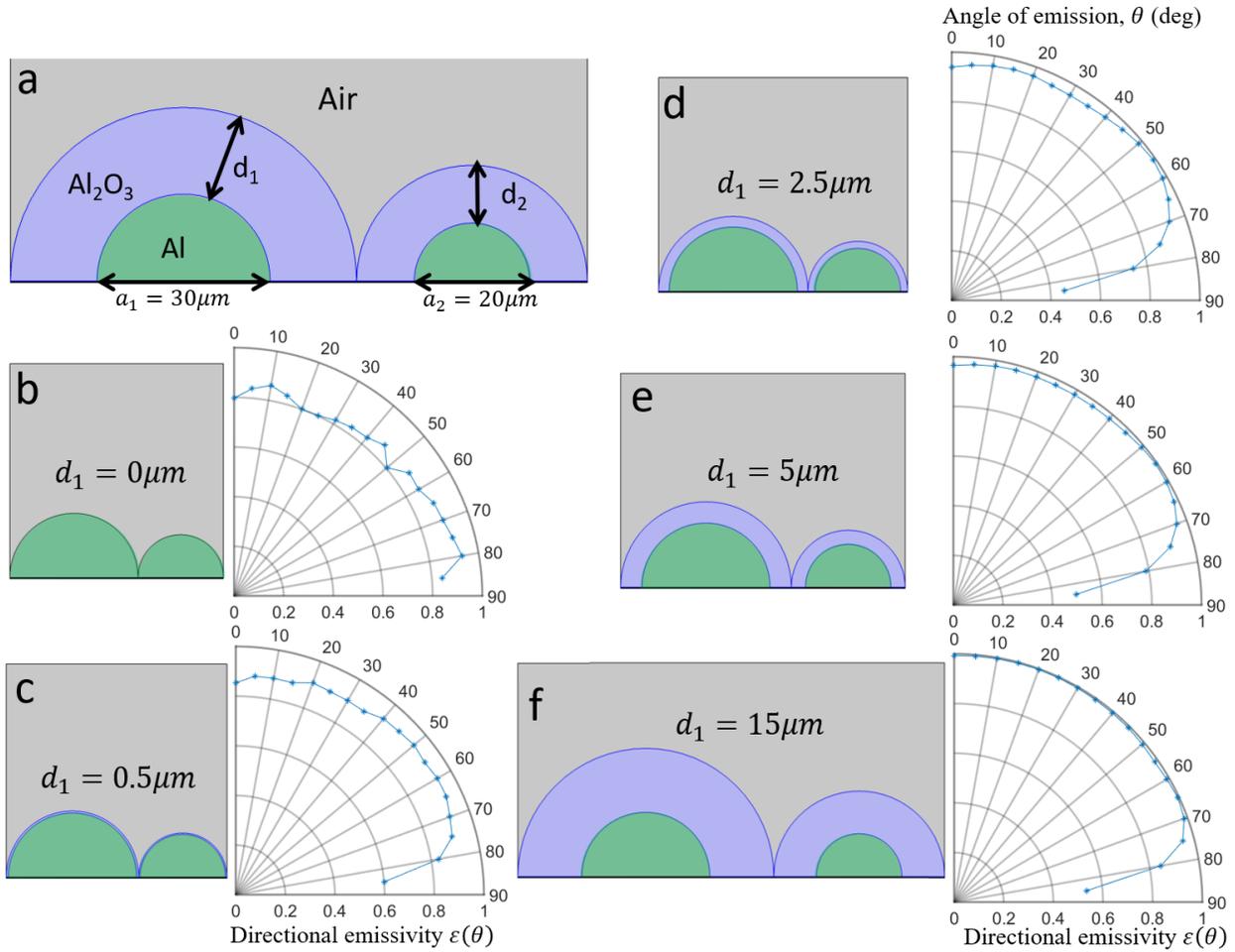

**Fig. S13 The emissivity of two periodic aluminum mounds with varying aluminum oxide thicknesses.** (**a**) Schematic of two periodic aluminum mounds with varying aluminum oxide thicknesses. The aluminum mounds have fixed diameters of $a_1 = 30$ μm and $a_2 = 20$ μm, respectively. (**b-f**) Left panels: geometries corresponding to different thicknesses of the aluminum oxide layer ranging from $d_1 = 0$ μm in (**b**) to $d_1 = 15$ μm in (**f**). The thickness of the second mound aluminum oxide layer $d_2$ changes proportionally to the thickness of the first oxide layer $d_1$ by using the relation: $d_2 = a_2 \cdot d_1/a_1$. Right panels: corresponding computed directional emissivity values.

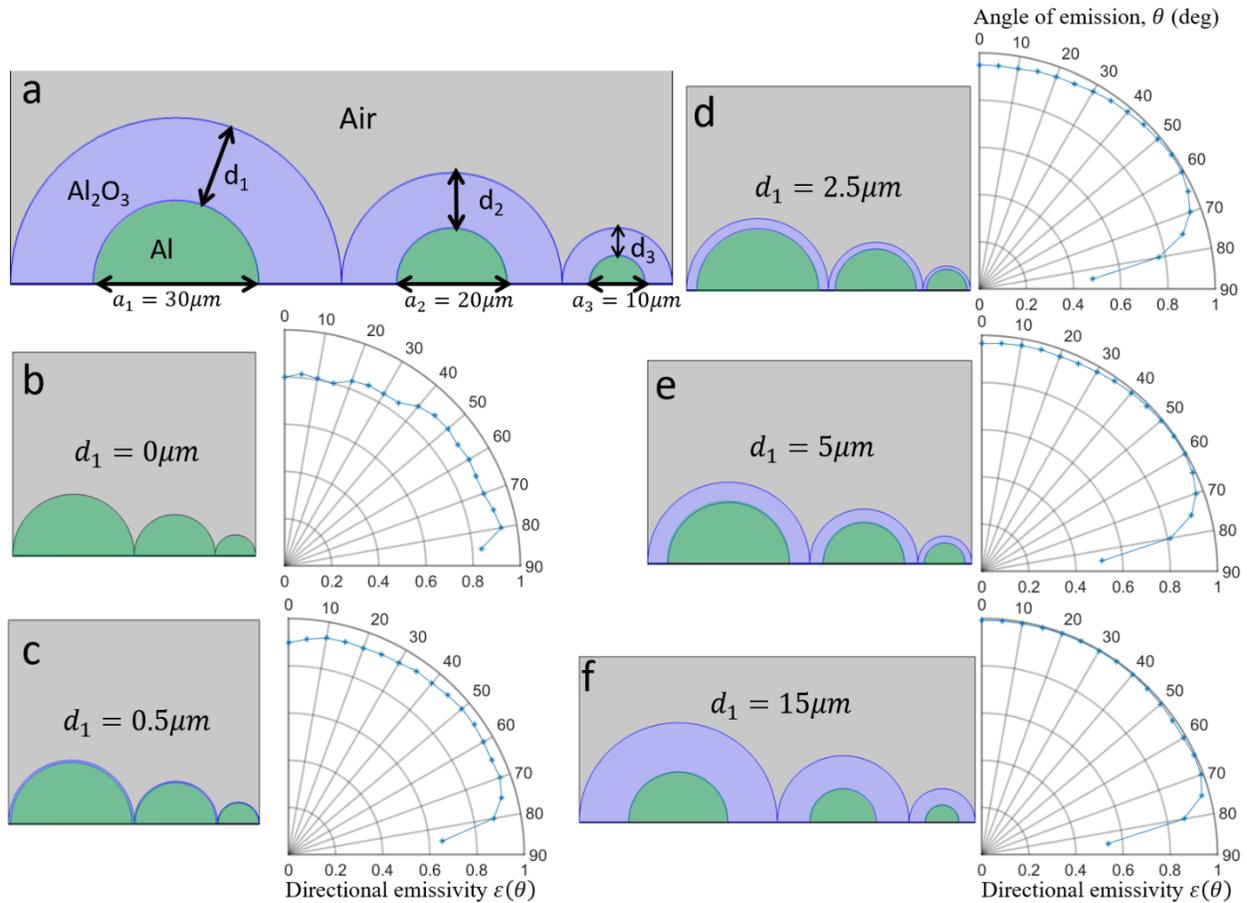

**Fig. S14 The emissivity of three periodic aluminum mounds with varying aluminum oxide thicknesses.** (**a**) Schematic of periodic three aluminum mounds with varying aluminum oxide thicknesses. The aluminum mounds have fixed diameters of $a_1 = 30$ μm, $a_2 = 20$ μm, and $a_3 = 10$ μm, respectively. (**b-f**) Left panels: geometries corresponding to different thicknesses of aluminum oxide layer ranging from $d_1 = 0$ μm in (**b**) to $d_1 = 15$ μm in (**f**). The thicknesses of the second and third mound aluminum oxide layers $d_2$ and $d_3$, respectively, change proportionally to the thickness of the first oxide layer $d_1$ by using the relations: $d_2 = a_2 \cdot d_1/a_1$ and $d_3 = a_3 \cdot d_1/a_1$. Right panels: corresponding computed directional emissivity values.

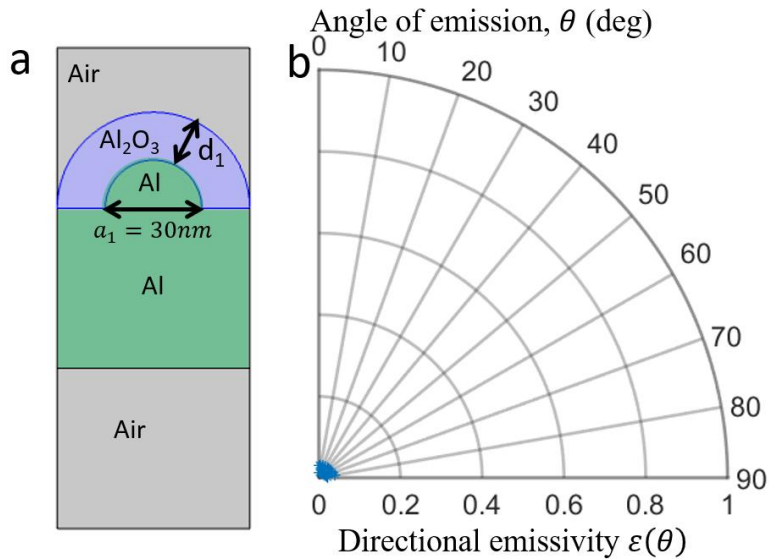

**Fig. S15 The resulted mid-IR emissivity is extremely low for nanoscale mound morphologies.** (**a**) Schematic of a single periodic aluminum mound with nanoscale dimensions ($a_1 = 30$ nm and $d_1 = 15$ nm). (**b**) Computed directional emissivity values. Note the directional emissivity for Fig. S2 is plotted from 0 to 0.15 but is plotted from 0 to 1 here.

## 7. Statistical information

The uncertainty in hemispherical emissivity values measured using the thermal imaging camera (TIC) based technique described in the Supplementary information is accounted to two causes. First, for most samples, because the directional emissivity is consistent across all angles with only a slight decrease after 65 degrees, the geometric error caused by approximating the value of the integral to calculate hemispherical emissivity from the directional measurements is less than 2%. The remaining uncertainty is accounted for in the 2% error from the thermal camera, as stated by the manufacturer. The reflection-based instrument (Surface Optics SOC-100) has an uncertainty of 1% overall for hemispherical and directional emissivity measurements, as quoted by the manufacturer. The maximum hemispherical emissivity value reported is the average of 24 measurements, two per each of the 12 samples produced in six batches using two laser systems at three different times. The standard deviation of the 24 measurements is also reported.

The reported measured surface roughness parameters are the average and standard deviation from the LSCM scans. For the background gas experiment, three LSCM scans from different locations on each sample were used. For the acid etching experiment, four LSCM scans were used, two scans per sample at each given etching parameter. For each scanned area, the average roughness ($R_a$) was measured over the entire scanned area. The average height was the average of the maximum height ($R_z$) measurement for ten subset areas within each scanned area.